\documentclass[aps,physrev,graphicx,amsmath,amssymb,reprint,superscriptaddress,nofootinbib]{revtex4-2} 
\usepackage{graphicx}
\usepackage{subfig}
\usepackage{mathtools}

\usepackage{dcolumn}
\newcolumntype{d}[1]{D{.}{\cdot}{#1} }
\usepackage{bm}

\usepackage{xcolor,soul}
\usepackage{physics}
\usepackage{tikz}
\usetikzlibrary{decorations.markings}

\usepackage{mathpazo} 

\usepackage{siunitx}

\draft 

\newcommand{\nb}[1]{\textcolor{purple}{#1}}

\begin{document}
\title{Learning the Fermion sign structure in path-integral Monte Carlo}
\date{\today} 

\author{Jarvist Moore Frost}
\affiliation{Department of Chemistry, Imperial College London, Exhibition Road, London  SW7 2AZ, UK}
\affiliation{Department of Physics, Imperial College London, Exhibition Road, London  SW7 2AZ, UK}
\email[Electronic mail:]{jarvist.frost@imperial.ac.uk}

\keywords{path-integral, Monte Carlo}

\begin{abstract}
Starting from a \emph{probabilistic numerics} approach to the Fermion sign problem in path integral Monte Carlo, we recast the arithmetic calculation of a Fermionic observable as a statistical inference problem. 
We develop approaches that learn the behaviour of Fermion exchange cycles binned by the conjugacy class of the permutation group (which we term `permutation family'). 
This extends the work of DuBois, Brown and Alder\cite{dubois2017overcoming} to inhomogeneous and more complex systems. 
Monte Carlo samples are used to train models for both the probability of a permutation family and the energy of this set of exchange permutations. 
The overall Fermionic energy is then directly inferred from these models, without using a direct ratio estimator on the Monte Carlo samples. 

By imposing physical understanding as inductive priors, we produce accurate and useful fits that remain robust even in regimes with severe sign problems. 
We generalise the linear (ideal-gas style) models of DuBois et al. with Bayesian priors that enforce the intuitive models of Feynman\cite{Feynman1953A} at their asymptotic limits. 
These linear models serve as the baseline for a Long Short-Term Memory (LSTM) neural network,  which is tasked with learning only the residual many-body \emph{correlations} on top of the physical model. 
We develop active important sampling methods driven by these models, which direct the Monte Carlo chains toward undersampled permutation regions, to efficiently reduce the variance in the observable. 

We apply this framework to small experiments on benchmark systems: the spin-polarised uniform electron gas, and electrons in a 2D harmonic confining potential. 
In both cases we demonstrate that this inference-based framework can extract stable energies in regimes where direct Monte Carlo sampling fails due to the sign problem. 

\end{abstract}

\maketitle

\section{Introduction}\label{introduction} 

\subsection{The Fermion sign problem in path-integral Monte Carlo}

In Fermionic path-integral Monte Carlo, we undertake a Bosonic simulation\cite{Ceperley1995} (where probabilities are well defined and positive), and then reweight the simulation picking up a factor of $(-1)^{p_{ex}}$ where $p_{ex}$ is the number of exchanges in the permutation $[p]$ with $N$ particles of which there are $N!$. 
For energy,

\begin{equation}
    \langle \hat{E} \rangle_F
    = \frac{\sum_{[p]} (-1)^{p_{\mathrm{ex}}} \, P_{[p]} \, E_{[p]}}
           {\sum_{[p]} (-1)^{p_{\mathrm{ex}}} \, P_{[p]}}
    \equiv \frac{\langle \sigma E \rangle_B}{\langle \sigma \rangle_B}.
\end{equation}

This ratio estimator has numeric issues when the denominator, the average sign sampled in the simulation $\sum_{[p]} (-1)^{p_{ex}} \, P_{[p]}$, approaches zero, as any sampling error is magnified by the division. Once the error in the denominator is comparable to its average value, the estimator fails entirely and gives wildly oscillating estimates. 

Unfortunately, as N increases (providing a larger and therefore more finely balanced set of permutations $[p]$), as temperature decreases (the particle exchange energy driving the system into longer permutation cycles), or the coupling increases (increasing particle exchange energy) the average sign is relentlessly pushed to zero. 

At this point the simulation has not failed: nothing untoward has happened in the autocorrelation time or ergodicity of the simulation. 
The problem is in our Fermionic reweighting, where the ratio estimator requires a finesse of accuracy that cannot be achieved with the usual $\sigma_s = \sigma_g / \sqrt{M}$ behaviour of random Monte Carlo samples. 

Our tenet is that though the Fermionic sign problem is NP-hard in the worst case scenario\cite{Troyer2005}, often what we observe practically as the exponential variance barrier in many physical Hamiltonians is not this insurmountable limit. 
Rather, the practical collapse is a direct consequence of the na\"ive arithmetic mean we are using as our estimator. 
Because we might hope that non-pathological physical systems possess learnable structure and correlations, we intend to bypass the \textit{statistical} inefficiency with a more sophisticated estimator. 
To this end we take a Probabilistic Numerics\cite{Hennig2015} perspective, and attempt to recast the evaluation of the Fermionic observable as a probabilistic \emph{inference} problem (what Fermionic observable would best explain these data) rather than a direct arithmetic calculation. 

\subsection{Permutation families}

To undertake inference on the path integral Monte-Carlo data samples we must first make the problem tractable---there are $N!$ raw permutations $[p]$, so directly learning this would then clearly have the same factorial scaling as the sign-problem itself. 

To this end, and throughout this paper and associated codes, we therefore aggregate individual permutations $[p]$ by their conjugacy class of the $S_N$ symmetric group. 
These are isomorphic to the integer partitions, and describe the different ways that $N$ particles can be decomposed into cycles. (For $N=2$ there are $2$: $1+1$ and $2$; for $N=3$ there are $3$: $1+1+1$, $1+2$ and $3$.)

For computational convenience, it was found that these were most easily described by a cycle-count vector\footnote{Initially we used the $\lambda_k$ notation of combinatorics, which is constructed by a slightly different counting constraint $\sum_{i=1}^N \lambda_i = N$, but found this more awkward.}, defined as 

\begin{equation}
    \bm{C} = (C_1, C_2, \dots, C_N), \quad C_\ell\ge0, \quad \sum_{\ell=1}^N \ell \, C_\ell = N .
\end{equation}

Here $C_\ell$ is the \emph{number} of cycles each of which has length $\ell$. 
As the particles are identical, there are a large number of equivalent ways to label the conjugacy class. 
We describe a given $\bm{C}$ as a `permutation-family' (previous authors used `permutation sector' or `conjugacy class', but as we use the Worm algorithm we already have $G$ and $Z$ `sectors' to consider, and we cannot spell or pronounce `conjugacy' reliably).

We can calculate this degeneracy (which is also the infinite temperature occupancy of the states) with combinatorics, 

\begin{equation}
    M(\bm{C}) = \frac{N!}{\prod_{\ell=1}^N \ell^{C_\ell} \, C_\ell!}.
\end{equation}

Perhaps surprisingly, there is no closed-form solution for the number of $\bm{C}$ for given $N$, and this must instead be calculated recursively. 
However, the Ramanujan-Hardy asymptotic formula $\sim \exp \left( \pi \sqrt{2N / 3} \right) / (4N\sqrt{3})$ suggests that the number of permutation-families that must be tracked expands sub-exponentially in $N$. 

A direct evaluation shows that $p(N=60)\approx10^6$ (therefore a dense histogram data structure of order megabytes) and $p(N=115)\approx10^9$ (of order gigabytes, reaching current reasonable limits on a single computational workstation). 
Therefore in all our current experiments we constructed a dense histogram data object. 
Care however was taken to ensure that the methods developed below should still work where $p(N)$ cannot be instantiated in memory, and instead a sparse key-value method of compiling data on the \textit{visited} sectors is used, which could be efficiently maintained on disk on a distributed network across multiple nodes, etc. 
The Monte Carlo algorithm will visit some of these sectors extremely rarely (i.e. not in finite time). 
The models we develop below can predict sensible probabilities even in these unvisited zero-data regions, by the imposition of physical priors which are then taken through the linear models and into the machine learning. 

As only the labels are shifting, all $[p]$ which are in a given $\bm{C}$ are formally identical. 
In particular, this can be inferred in a simulation by calculating a canonical representation of $\bm{C}$ for a given simulation $[p]$: as these samples are formally identically, we are strongly motivated to aggregate them (i.e. histogram based on $\bm{C}$). 
The Fermion sign within a given $\bm{C}$, by counting the total cycle count $n_c = \sum_\ell C_\ell$, is $S_C=(-1)^{N-n_c}$. 
Hence, inside any given permutation family, the sign is constant and the probability is entirely well-defined.

We therefore intend to evaluate our estimators within these permutation-families, where the sign is fixed.

In 2014, Du Bois, Brown and Alder\cite{dubois2017overcoming, brown2014thesis-PATHINTEGRALMONTE} (hereafter referred to as `DBA') investigated the path-integral Monte Carlo of these conjugacy classes in homogenous systems. 
Following Feynman\cite{Feynman1953A, Feynman1953B} who seems to have first considered these structures in his theory of the $\lambda$ transition in helium, they infer the probability of a pair exchange $p_2$ from the cycle statistics, and model the full sector distribution using a single exchange-penalty parameter $\kappa$. 
Overall they find fantastic reducion of error in two key homogenous systems: the spin polarised homogenous electron gas (at $N=33$, $r_s=\{1,10\}$, $\Theta=\{1,1/8\} \quad T_F$), and Fermionic $N=66$ $^3\text{He}$. 
Future work was to consider \emph{heterogeneous} systems, but this does not seem to have been published.

More recently, Dornheim et al.\cite{Dornheim2019permutation} looked at the 2nd-moment of the distribution of permutation cycle lengths for a wide range of systems in a well-founded and validated path-integral Monte Carlo code. 
They find that the independence of cycle length found in DBA does not hold for these systems: there is clear statistical dependence between cycle lengths.  

Our aim in this work is to continue the DBA approach, and build models which contain sufficient flexibility to model complex systems. 

Our central tenet is that we should reinterpret the goal of quantum Monte Carlo with a probabilistic numerics lens, and see the calculation of quantum observables (here, energy throughout) as an inference problem. 
From this point of view, the Fermion sign-problem becomes a matter of fitting models with suitable finesse, something that can be done within permutation families (where all of the samples have the same Fermion sign) without particular numeric difficulties, where probabilities and estimators can be well defined. 

Therefore the aim of this paper is exploratory and constructive: to develop automatic and statistically motivated approaches to fit models directly from path-integral Monte Carlo data, including the well motivated physical inductive biases used by Feynman and DBA, as well as modern deep-learning approaches to learn non-linear and high dimension correlations. 

Considerable recent success has been made with variants of Neural Network Quantum States\cite{Hermann2023}. 
These high-dimension wavefunctions are typically learnt by sampling local energy gradients, an incredibly lean data-source considering how rich the representation needs to be. 
In contrast, the permutation-family approach presented here provides a richer structured data source directly derived from Monte Carlo samples (i.e. the actual phase-space state vectors), combined with physically motivated priors (exact cycle multiplicities and ideal-gas limits), which both feed and constrain the statistical linear and deep-learning models. 

\section{Method}\label{method}

Path-integral Monte Carlo (PIMC) discretises the imaginary-time propagator into $M$ time slices of duration $\tau = \beta/M$, so that each particle is represented by a ring polymer of $M$ beads.
We use path integral Monte Carlo with the continuous-space Worm algorithm\cite{Boninsegni2006A,Boninsegni2006B}
with the periodic boundary corrections of Spada et al.\cite{Spada2022}. 
We implemented the methods in this paper in a green-field code, following Spada et al.\cite{Spada2022} closely in the Bosonic implementation, and then adding Fermion re-weighting (including via the Xiong and Xiong\cite{Xiong2022} $-1<\zeta<1$ fictitious sign approach, though we do not discuss that approach further).

To store and manipulate the $p(N)$ family-permutations, we need an approach to label the cycle-count vectors, ideally as a dense set of integer ranks $k \in \{1, \ldots, p(N)\}$. 
Eventually we came up with an optimised approach using the \emph{integer partition count table} $P[n,m]$, defined by recurrence
\begin{equation}\label{eq:partition-table}
\begin{aligned}
      P[n,m] &= P[n, m-1] + P[n-m, m],\\
  P[0,m] &= 1 \quad \text{ for } m>=1,\\
  P[n,0] &= 0 \quad \text{ for } n > 0,
\end{aligned}
\end{equation}
where $P[n,m]$ counts the number of integer partitions of $n$ with all parts $\leq m$.
In particular, and as a key check, the diagonal $P[N,N] = p(N)$.
The table is computed once in $\mathcal{O}(N^2)$ time by dynamic programming.

Given a $\bm{C}$, we convert to a descending integer partition and rank it against $P$ in $\mathcal{O}(N)$ time by walking down the table and accruing the `skip' value; the inverse (unranking) is likewise $\mathcal{O}(N)$ as you constructively walk up the table. 
This permits fast lookup of permutation-family rank and so update accumulators during the Monte Carlo.

\subsection{Models of permutation}\label{permutation-model}

Following Feynman and DBA, we realised that we could define a general class of what we termed (exponential) $\theta$ models where each individual permutation cycle $\ell$ has an independent penalty applied against it, $\exp (\theta_l C_\ell)$, modifying the raw multiplicity (degeneracy) $M(\bm{C})$,

\begin{equation}
    Q_m(\bm{C}) = \frac{1}{\mathcal{Z}(\bm{\theta})} \, M(\bm{C}) \exp \left( \sum_{\ell=1}^N \theta_\ell \, C_\ell \right).
\end{equation}

The logarithm $\log M(\bm{C})$ serves as the entropic baseline: in the high-temperature limit ($\beta \to 0$), where exchange penalties vanish, the sector probability is proportional to $M(\bm{C})$.

\subsection{Fitting these models}

These models impose an assumption that the probability of a given cycle length is entirely independent of the other cycles (it is essentially a fugacity). Therefore it is a property of the average cycles $\hat{p}_\ell = \langle C_\ell \rangle$, and so we can aggregate across all permutation-family samples from Monte Carlo. 
In this framework, the physical free-energy cost of forming the $\theta_l C_\ell$ cycle is directly the log-probability prior expectation. 

To fit these models, we minimize the Negative Log-Likelihood (NLL) of the model distribution $Q_m$ evaluated against the empirical Monte Carlo sector probabilities $\hat{P}(\bm{C})$.

The total NLL across the full distribution is  

\begin{equation} \mathcal{L}_{\mathrm{NLL}} = -\sum_{\bm{C}} \hat{P}(\bm{C}) \cdot \log Q_m(\bm{C}), \end{equation}

where we are summing over all permutation-family vectors $\bm{C}$.

Substituting our exponential family of models $Q_m(\bm{C})$ into this loss,

\begin{equation} \mathcal{L}_{\mathrm{NLL}} = -\sum_{\bm{C}} \hat{P}(\bm{C}) \cdot \left[ \log M(\bm{C}) + \sum_{\ell=1}^N \theta_\ell C_\ell - \log \mathcal{Z}(\bm{\theta}) \right] \end{equation}

Because the empirical probabilities sum to 1 ($\sum_{\bm{C}} \hat{P}(\bm{C}) = 1$), the $\log \mathcal{Z}(\bm{\theta})$ term factors out. 
Further, the empirical expectation of the cycle count is given by $\langle C_\ell \rangle = \sum_{\bm{C}} \hat{P}(\bm{C}) C_\ell$, which we denote as $\hat{p}_\ell$.

Dropping the constant relative entropic baseline $-\sum \hat{p}(\bm{C}) \log M(\bm{C})$ (which does not depend on $\bm{\theta}$), our effective loss function simplifies to just the parameters and their empirical conjugate expectations: 

\begin{equation} \mathcal{L}_{\mathrm{eff}}(\bm{\theta}) = \log \mathcal{Z}(\bm{\theta}) - \sum_{\ell=1}^N \hat{p}_\ell \theta_\ell \end{equation}

This simple to calculate form naturally gives greater weight to well-visited permutation-families (by $\hat{p}_\ell$), has only $N$ parameters, and is well behaved in optimisation.

\subsubsection{Infinite temperature classical limit ($\theta_\ell = 0$)}

In the classical, infinite-temperature, limit where the pair-exchange energy is trivial compared to the thermal energy, the system fully samples the permutation-families with the weight of their multiplicity (statistical microstates). 
Imposing all $\theta_\ell = 0$ entirely takes $\exp(0) = 1.0$, the probability model cleanly falls onto $Q_0(\bm{C}) \propto M(\bm{C})$, the purely entropic baseline.

\subsubsection{Feynman $p_2$ (pair-exchange) model ($\theta_\ell = -\kappa (\ell - 1)$)}

Following Feynman and DBA, macroscopic exchanges can be constructed from a penalty $\kappa > 0$ applied per pairwise exchange. 
Imposing 
\begin{equation}
\begin{aligned}
    \theta_\ell &= -\kappa (\ell - 1) 
    \quad \implies \quad \\
    Q_\kappa(\bm{C}) &= \frac{1}{\mathcal{Z}(\kappa)} \, M(\bm{C}) \exp \left( -\kappa \cdot n_{\mathrm{swaps}} \right),
\end{aligned}
\end{equation}
where $n_{\mathrm{swaps}} = N - \sum_{\ell} C_\ell$ is the number of pair-wise swaps required to generate the cycle. 
A simple Nelder-Mead gradient-free (blackbox) optimisation procedure was used, to not require calculating the gradient of the above.  

\subsubsection{Maximum entropy $\theta_\ell$ model}

To try and consider situations where cycle-length correlations are present, we fit the full $N-1$ free parameters $\theta_2, \ldots, \theta_N$ (with $\theta_1 \equiv 0$ by convention).
This is a maximum-entropy distribution subject to matching the empirical cycle-count expectations $\langle C_\ell \rangle$.

To attempt to reduce over-fitting, we impose weak isotropic $L_2$ regularisation, 
\begin{equation}\label{eq:maxent-loss}
  \mathcal{L}_\mathrm{MaxEnt}(\bm{\theta})
  = \log\mathcal{Z}(\bm{\theta}) - \sum_\ell \hat{p}_\ell \theta_\ell
  + \lambda_\mathrm{reg}\sum_{\ell=2}^{N} \theta_\ell^2,
\end{equation}
where $\lambda_\mathrm{reg}$ is a small regularisation strength (typically $10^{-4}$).

However, we found this approach still highly unstable: the $\theta_\ell$ coefficients seemed fairly random and wildly oscillated between Monte Carlo runs. 

\subsubsection{Maximum a posteriori}

Due to the failure of the unconstrained fit, we decided to use a maximum a posteriori (MAP) estimate with a baseline model (such as a $\kappa$ model as described previously).
We therefore decompose our fit as $\theta_\ell = \theta_\ell^\mathrm{prior} + \Delta_\ell$, and place a Gaussian prior $\Delta_\ell \sim \mathcal{N}(0, \tau_0^2)$ on the deviations, with $\tau_0^2 = 1.0$ in our experiments.

Our loss is now quite complex, but has a simple motivation 
\begin{equation}
\begin{aligned}
  \mathcal{L}_\mathrm{MAP}(\bm{\Delta})
  &= \log\mathcal{Z}(\bm{\theta}^\mathrm{prior} + \bm{\Delta})\\
  &- \sum_\ell \hat{p}_\ell (\theta_\ell^\mathrm{prior} + \Delta_\ell)
  + \frac{1}{2\tau_0^2}\sum_{\ell=2}^{N} \Delta_\ell^2.
\end{aligned}
\end{equation}

Where data is plentiful (such as with short cycle lengths), the data dominate and the model can pull away from the $\kappa$ baseline. 
Where observations are sparse, the prior smoothly pulls $\Delta_\ell \to 0$ and recovers the baseline.

To quantify the fidelity of the analytical prior, we compute the Kullback-Leibler divergence from the model distribution $Q_{\text{MAP}}$ to the empirical Monte Carlo distribution $\hat{P}$, defined as $\text{KL}_{\text{MAP}} \equiv D_{\text{KL}}(\hat{P} \parallel Q_{\text{MAP}}) = \sum_{k} \hat{P}(k) \ln \left( \frac{\hat{P}(k)}{Q_{\text{MAP}}(k)} \right)$, where the discrete summation runs over all permutation sectors $k$ with non-zero empirical probability.

\subsubsection{Long short-term memory (LSTM)}

The word-like structure of the permutation family C-vector naturally suggests an autoregressive machine-learning model, where the $\bm{C}$ vector is represented as an ordered sequence of individual cycle lengths, $S = (s_1, s_2, \ldots, s_K)$. 
Clearly the probability of filling the $k$'th element of the $C_k$ vector should depend on the preceding terms. 
The celebrated long short-term memory\cite{Hochreiter1997} (LSTM) is a natural probabilistic model for the element-by-element construction of these vectors, updating an internal memory as it progresses through the sequence. 
Rather than autoregressive across time, the LSTM is learning how to allocate a finite particle pool into progressively longer exchange topologies. 
The representation at each stage of the pseudo-word is one of probability: a distribution over possible choices in the count ($x_l$) for the next cycle $\ell$ under consideration. 
The (telescoped) probability at the end of having read through the full sequence is exactly our probability for this permutation-family. 

These are expressed as logits (log-odds), which map $p=(0,1)$ onto an unnormalised $(-\infty, \infty)$, transformed back into strict probabilities with a `Softmax' $ p_c = \frac{\exp(x_c)}{\sum_j \exp(x_j)} $. 
Within the $C_k$ vector representation we found that we could enforce exact conservation of particle number very naturally, by taking our exact known prior probability given the remaining `mass' of particles, and then simply add the logarithm to the logits before the `Softmax'. 
Any choices which would exceed the number of particles left ($p=0$) are therefore pushed to $-\infty$ before the `Softmax' normalisation, strictly conserving mass while ensuring the remaining probability is well defined (sums to one). 
No expressive power of the model is wasted in learning where the boundaries of possibility are.

Additionally this means that any $\theta$ exponential family of probability models can be naturally used as the baseline around which the model trains. 
These are brought in by evaluating the (unnormalised) free statistical weight (fugacity) of adding a single cycle of length $\ell$ as $ f_\ell = \frac{1}{\ell} \exp(\theta_\ell) $.
The initial machine-learning model (with final layer weights initialised at zero) makes no correction, and so exactly reproduces the $\theta$ exponential probability model prior. 

In sum therefore, the per-character ($C_\ell$) probability is 

\begin{equation}
    P(C_\ell = c ) \\
    \propto \exp\!\Big( x_{\ell,c}
    + \ln \big[ P_{\mathrm{prior}}\big(C_\ell = c \mid N_{\mathrm{rem}}\big) \big] \Big),
\end{equation}

where $x_{\ell,c}$ is the logit for assigning count $c$ to cycle length $\ell$, and $N_{\mathrm{rem}} = N - \sum_{j<\ell} j C_j$ are the number of remaining particles to assign. 
The prior assigns $P_{prior} = 0$ (logit ($-\infty$) to any $c$ with $\ell c > N_{\mathrm{rem}}$, as this can no longer be constructed with the remaining particles. 

Our initial experiment used an unbiased LSTM to fit the probability of that permutation-family. 
Unsurprisingly, this led to the same trap as a na\"ive $\theta_\ell$ model: permutation-families which had not been observed in the Monte-Carlo were assumed to be just-below the noise floor (for what else should the model expect?).

\subsection{Models of energy}\label{energy-model}

The na\"ive global Monte Carlo estimator for the (Bosonic) system energy is the weighted sum of the permutation-family arithmetic means, 

\begin{equation} E_{\text{tot}} = \sum_{k} \hat{p}_k \langle E \rangle_k \quad 
\text{where} 
\quad \langle E \rangle_k = \frac{1}{n_k} \sum_{i=1}^{n_k} E_{k,i}. 
\end{equation}

Here $n_k$ is the estimator count in that permutation-family, and $E_{k,i}$ are the individual estimates. (In our code we collate $\sum_{i=1}^{n_k} E_{k,i}$ as we go along.)

Partitioning the estimators over permutation-families yields the exact same statistics as if we had globally summed over these values, but crucially provides a basis for inference. 

In the previous section we developed models for the probability of permutation-families which could include physical understanding as inductive bias. 
This helps with the variance that would otherwise be present in $n_k$, but does nothing to help with the sample variance in $E_k$. 
To make some progress at alleviating the Fermion sign problem, we need to fit similar physically justified models for the energy which can then be summed against our models of permutation-family probability. 

\subsubsection{Linear energy model}

Making the assumption that the total energy of a permutation-family is the sum of the individual $\ell$ loops (i.e. ignoring interactions between specific permutation cycles), we can decompose the energy into a linear sum of the loop count numbers. Without loss of generality, we can restate this in the suggestive form, 

\begin{equation}
    E_\ell = \ell \cdot E_{MF} + \Delta_\ell ; 
    \quad \text{with} \quad
    \Delta_1=0 .
\end{equation}

The constraint $\Delta_1=0$ fixes the gauge and the collinearity of the fit, extracting $\ell\cdot E_{MF}$ as a mean-field energy (per particle, independent of the size of the exchange cycle). 

We solve for $\{E_{\text{MF}}, \Delta_2, \dots, \Delta_N\}$ using Weighted Least Squares (WLS) on the Monte Carlo sector energy estimates $\langle E \rangle_{[p]}$, subject to two (Bayesian) priors controlled by hyperparameters $\lambda_{\text{ridge\_}\Delta}$ (a ridge penalty applied to the residuals $\Delta_\ell$) and $\lambda_{\text{smooth}}$ (a smoothness prior). 
Initially priors were introduced due to observed oscillations (noise) in the fit, but were then adapted and adjusted to provide a statistically motivated approach to reproduce the two limiting cases that DBA explored.

\begin{figure}
    \centering
    \includegraphics[width=1.0\linewidth]{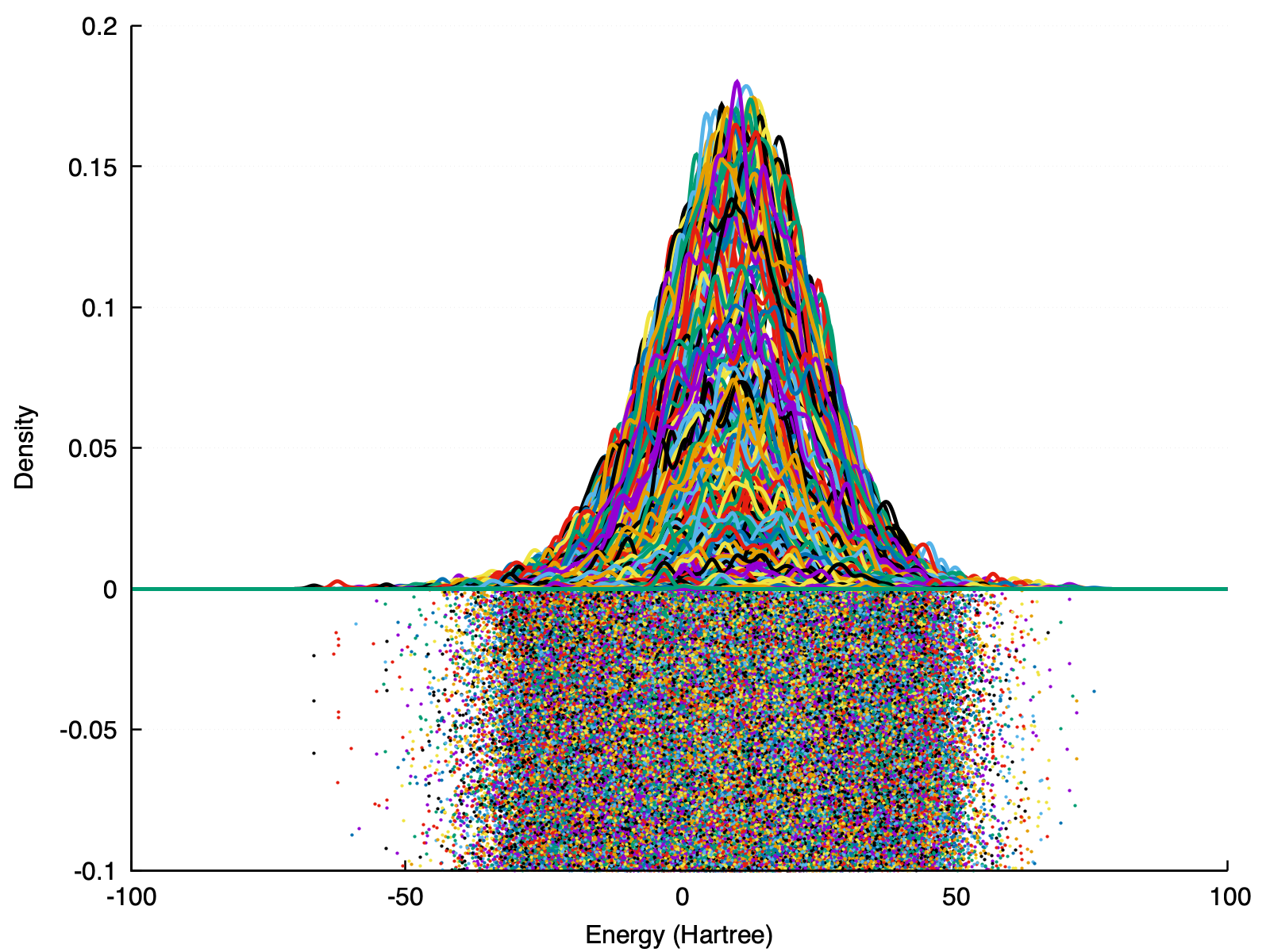}
    \includegraphics[width=1.0\linewidth]{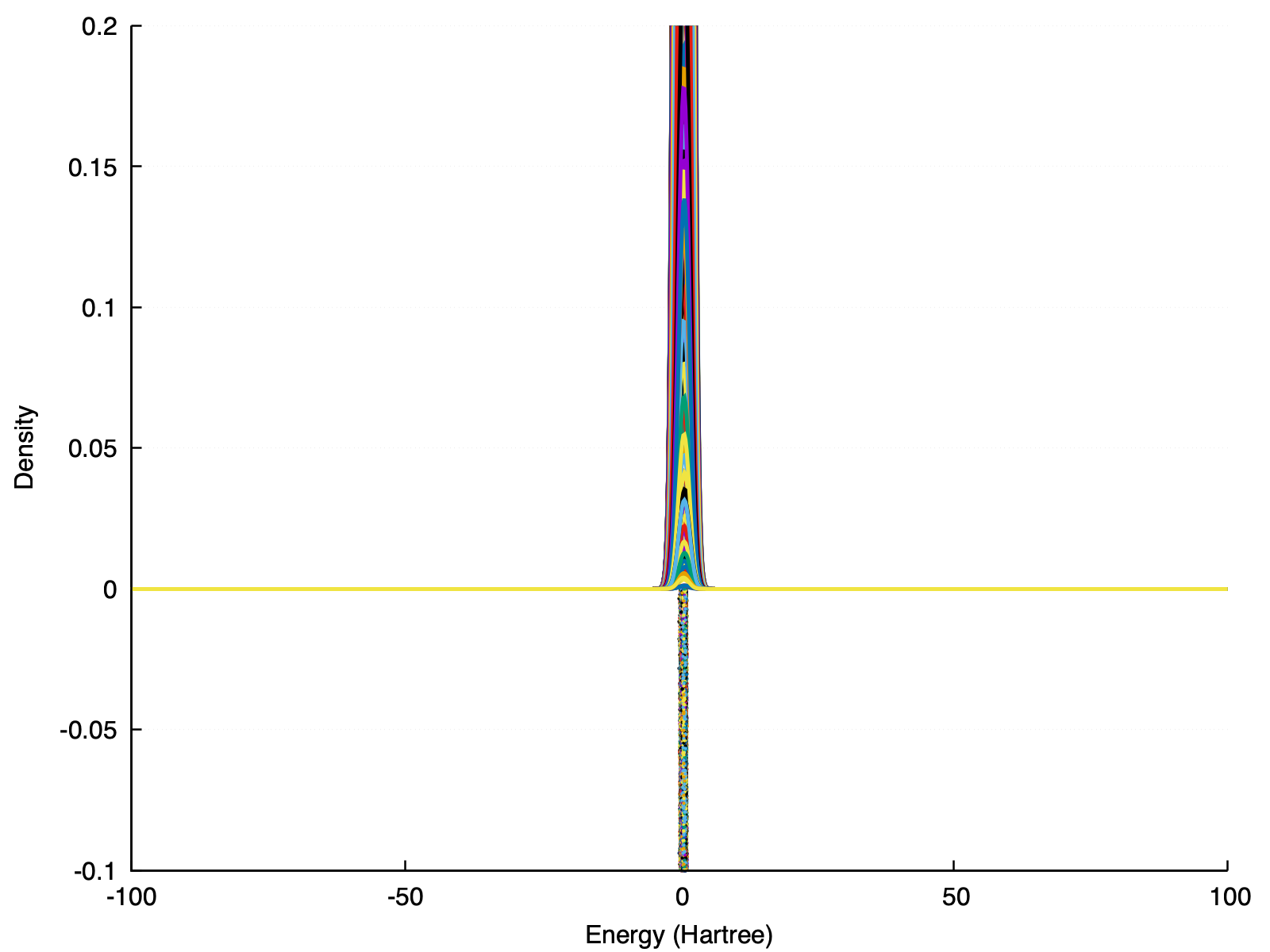}
    
    \caption{Up to 1000 energy samples taken from each permutation-family in an $N=19$ spin-polarised UEG simulation, with the Yakub-Ronchi potential and Kelbg smoothing. 
    Temperature in both figures is `low' ($\Theta=0.125$). 
    Top is high-density $r_s=1.0$, bottom is low-density $r_s=10.0$. 
    The higher density clearly forces much larger variance in the energy. 
    On the bottom is a `rug' plot with the different sectors different colours, and the datapoints jittered (offset randomly in the y-axis) for visibility. 
    On the upper, with the same colour, are kernel density estimators over the same data with a bandwidth of 1.0 (Hartree). 
    Typically the low ranked cycles (fewer permutations) seem to have a smaller variance. 
    A vastly larger variance of energy is sampled in the high density regime---individual permutation-family samples can have very different energy. 
    Yet, the data permutation-family samples look evenly distributed and fairly Gaussian
    }
    \label{fig:ueg_reservoir}
\end{figure}

Figure~\ref{fig:ueg_reservoir} shows energy samples from the permutation family: the energy distributions are broad and temperature dependent, but reassuringly Gaussian and with an variance which seems to increase strongly with density. 

\subsubsection{The ideal-gas limit}

The ridge penalty, applied strictly to the residuals, attempts to minimise the size of the linear components of the fit. 
Taking $\lambda_{\text{ridge\_}\Delta} \to \infty$ therefore forces all $\Delta_\ell$ to zero, leaving us with the fit entirely carried by $E_{MF}$,

\begin{equation}
    E_{\ell} = \ell \cdot E_{\text{MF}} \quad \Rightarrow \quad E_{\text{tot}} = N \cdot E_{\text{MF}}.
\end{equation}

This is the ideal-gas limit, where energy is simply proportional to the number of particles, independent of the exchange cycle or permutation-family (indeed, any interparticle interaction). 
DBA suggested this was their most robust estimator at high temperature ($T \gg T_F$) where quantum effects are weak, and many permutation-families are visited leading to low density of points in those families and poor quality statistics.

\subsubsection{The Feynman exchange penalty limit}

The smoothness prior penalises the second, discrete, derivative of the residuals. We achieve this by adding $\Delta_{\ell+1} - 2\Delta_{\ell} + \Delta_{\ell-1}$ to the Design matrix. 
In the limit of infinite $\lambda_{\text{smooth}}$, this forces the residuals to grow linearly with the length of the permutation cycle,

\begin{equation}
    \Delta_{\ell} = \kappa (\ell - 1).
\end{equation}

Substituting this into the total energy expression and simplifying, 
\begin{equation}
\begin{aligned}
    E_{\text{tot}} &= N E_{\text{MF}} + \kappa \sum_{\ell=1}^N C_{\ell} (\ell - 1) \\
    &= N E_{\text{MF}} + \kappa (N - n_{\text{cycles}})
\end{aligned}
\end{equation}
where $n_{\text{cycles}} = \sum C_{\ell}$ is the total number of cycles. 
Since $n_{\text{swaps}} = N - n_{\text{cycles}}$ is exactly the number of \emph{pairwise} exchange swaps required to generate the permutation, this is equal to 
\begin{equation}
    E_{\text{tot}} = N E_{\text{MF}} + \kappa \cdot n_{\text{swaps}}.
\end{equation}

Thereby we recover the Feynman exchange-penalty model developed in their $\lambda$ theory\cite{Feynman1953A,Feynman1953B}, where \emph{each} pair-exchange costs a fixed amount of action/energy $\kappa$ (a \emph{two} parameter model, with $E_{MF}$). 

\subsubsection{More complex linear models}

By choosing our 
$\lambda_{\text{ridge\_}\Delta}$
and $\lambda_{\text{smooth}}$, we can produce fits that apply varying degrees of these two physical priors in a clean Bayesian approach: where the data strongly disagrees with the inductive bias, it will make the largest changes. 

In order to then understand the resulting fits, we found it useful to evaluate the number of effective degrees of freedom ($d_{\text{eff}}$) of the model. 
In generalised ridge regression, this is given by the trace of the hat matrix, $\bf{H}$, 

\begin{equation} 
d_{\text{eff}} = \mathrm{Tr}\left[ (\mathbf{X}^T \mathbf{W} \mathbf{X} + \mathbf{P})^{-1} (\mathbf{X}^T \mathbf{W} \mathbf{X}) \right] 
\end{equation} 

where $\mathbf{W}$ is the diagonal weight matrix and 
$\mathbf{P} = \lambda_{\text{ridge}} \mathbf{I} + \lambda_{\text{smooth}}\mathbf{D}^T\mathbf{D}$ 
is the total regularisation penalty. 

\subsubsection{Neural network models}

The linear energy models capture the baseline behaviour, but intentionally ignore cycle-to-cycle interactions (where the energy cost of a specific permutation-family depends non-linearly on the other cycles). 
To capture these correlations, we build neural network models to capture this residual many-body \emph{correlation} energy, 
\begin{equation} E_{\text{NN}}(\bm{C}) = E_{\text{linear}}(\bm{C}) + \text{MLP}_{\bm{\phi}} \Big( \big[ \bm{h}_{\text{LSTM}}(\bm{C}) \oplus (\bm{C}/N) \big] \Big). 
\end{equation}

Here, $\text{MLP}_{\bm{\phi}}$ is a (shallow) Multi-Layer Perceptron (the ``energy head''), with trained weights $\bm{\phi}$\footnote{The permutation probabilities were modelled by a two-layer autoregressive LSTM (embedding dimension 16, hidden cell state 64). The residual energy head was a single-hidden-layer MLP (32 units, GELU activation). Both networks were trained via the Adam optimizer (learning rate $10^{-3}$, weight decay $10^{-3}$).}. 
The input to this model are formed by concatenating a scaled set of the permutation-family count vector $\bm{C}/N$ with the (frozen) final hidden state representation $\bm{h}_{\text{LSTM}}$ from the probability LSTM described in the previous section.

(In our first experiments we did not provide the MLP with the $\bm{C}/N$ vector; including it fairly unambiguously improved the predictive power of the model.)

We train against the residual energy left over after the linear model, i.e. $\Delta E_k = \langle E \rangle_k - E_{\text{linear}}(\bm{C}_k)$. 

To suppress the disruptive influence of noisy estimators in poorly visited permutation-families, having been slightly aghast after having visualised some of the raw estimates within permutation-families, we used a Huber loss with a transition at $1\,\text{Ha}$. 
The Huber loss grows quadratically for small residuals but transitions linearly for large errors, and is therefore less sensitive to extremes.

In some regions of the UEG parameter space, this still led to poor fits. 
We realised that the permutation-family sample variance estimators $s_k^2$ for low visit permutation-families ($n_k \le 10$) were unacceptably noisy and severely disrupted the fit if we included them na\"ively as the weight $W_k = n_k / \sigma^2$. 
At the same time, plotting the distribution of energies within each permutation-family showed that the true variance could vary considerable between permutation family. 
We therefore applied a standard statistical technique of empirical Bayes variance shrinkage: we mix in $\gamma$ pseudo-counts which carry the global variance (calculation across all permutation families). 

\begin{equation}
    \tilde{\sigma}_k^2 = \frac{n_k s_k^2 + \gamma_0 \, s_g^2}{n_k + \gamma_0};
    \qquad W_k = \frac{n_k}{\tilde{\sigma}_k^2}.
\end{equation}

Here $\gamma$ is a free, continuous, parameter which can be understood as the (number of) prior pseudo-counts used to anchor the variance, and $s_g^2$ is the global empirical sample variance. 
In our limited experiments we found that $\gamma=3$ was already sufficient to stabilise the linear fits.

\subsection{Sampling important permutations}\label{importance-sampling}

Unbiased Path-Integral Monte-Carlo will often fail to explore the permutation-family space. 
As the underlying simulation is Bosonic, at low temperatures it will enter a superfluid state where the vast majority of simulation time is spent in one or two extremely large permutations. 
In warm dense simulations, the opposite problem occurs and the Bosonic simulation will sit entirely in very small permutation cycles, as there is no Fermionic correlation hole to force otherwise. 

Upon reweighting to the Fermionic estimators, there is clearly very little statistical support. 

Therefore there is a strong motivation to apply importance sampling to the Monte Carlo to get more reasonable estimates in regions of space which are important to Fermionic matter. 

Originally developed in the context of calculating density of states in classical statistical mechanics, the Wang-Landau algorithm\cite{Wang2001} was developed to explore phase space and build a sampling-distribution inverse to the density of states. There have been many applications of the original idea, including to quantum Monte Carlo\cite{Inglis2013}.

Having built probability models to describe the permutation-family, we can directly use this information to bias the Monte Carlo and attempt to increase the Fermionic support. 

In our codes, following Spada et al.\cite{Spada2022}, the only Monte Carlo move which changes permutation family is the \textsc{swap!} move. 

To implement the bias, we consider the probability of the two permutation families which our move would be moving between, and then bias the acceptance probability to drive this into equal population, generating an overall flat histogram, in the style of Wang-Landau. 
If the probability model were perfect, the Metropolis algorithm would visit each permutation family evenly. Of course, our models are not perfect, nor do we necessarily want a fully flat histogram, and so we introduce a softening parameter $0 < \alpha < 1$. 
Indeed, applying the full $\alpha=1$ bias was found to tend to trap the Worm in a rather restricted subset of permutation-families, often entirely disjoint with the ones sampled in the unbiased Monte Carlo. 
In our limited experiments, we found that $\alpha=0.5--0.7$ was sufficient to drive the Monte-Carlo from highly localised (either very large superfluid exchange cycles; or very few exchange cycles) across more of the spectrum of permutation families. 

\begin{equation}
    A_{\mathrm{swap}}^{\mathrm{biased}} = A_{\mathrm{swap}} \times 
    \left( 
    \frac{Q_m(\bm{C}_{\mathrm{old}})}
        {Q_m(\bm{C}_{\mathrm{new}})} 
        \right)^{\alpha} 
\end{equation}

Practically we pre-compute $Q_m$ for each permutation-family, which enables us to use both the lightweight $\theta$ models, and also computationally much more expensive (of order miliseconds) LSTM-based probability models. 
Because both the analytic and LSTM-based probability modules are therefore frozen during the Monte Carlo, detailed balance is strictly preserved with respect to the modified target distribution.

Having run biased Monte Carlo, we then need to unbias the visiting statistics if we wish to use this information further. 
We have directly modified the distribution by $[Q_m(\bm{C})]^{-\alpha}$, we therefore debias by simply applying an inverse-probability weight element, defined per permutation-family, as $\mathcal{W}(\bm{C}) = [Q_m(\bm{C})]^{\alpha}$. 

\subsubsection{Variance reducing bias with the Jackknife estimator}

The Jackknife\cite{Miller1964,Miller1974} is an adaptable and easy-to-implement class of methods which can be used to reduce leading-order bias in a simulation (at a slight cost of increased variance; though in some ratio problems it has been shown that the variance can also be reduced\cite{Miller1964}), while also providing error estimates even for non-linear statistics. 
The method is based in `leave one out' statistics, which in in the block-Jackknife can be very naturally achieved by summing over the individual Monte-Carlo chains (each of which has run independently on a CPU core). 

By leaving out one entire Monte Carlo chain's data at a time, we iteratively compute Jackknife pseudo-values for the marginal permutation-family probabilities $p_k$ and the conditional (sign-free) permutation-family energy contribution $p_k \cdot E_k$. 
The sample variance of these (per permutation-family) pseudo-values gives us robust estimates of the standard-error.

These stratified (by permutation family) variance estimates we can directly use to drive importance sampling into the regions that most contribute to the overall energy: Neyman (optimum) allocation states that we want to sample proportional to the standard error in $p_k \cdot E_k$.

\subsection{Implementation}\label{implementation}

In order to implement these new methods, a fresh Path Integral Monte Carlo code, \textsc{Halcyon.jl}\footnote{By construction, all Worm-algorithm path integral codes\cite{Boninsegni2006A,Boninsegni2006B} must be named after sisters.}, was developed in the Julia programming language\cite{bezanson2017julia}. 
Use of Julia was key for the success of this work, as the same small codebase was used for initial implementations, machine-learning, statistics and plotting, and the deeply optimised Monte Carlo code itself. 

Spada et al.'s\cite{Spada2022} clear paper was used to implement the initial Bosonic code, which was then trivially extended to Fermionic resampling. 
For computational simplicity, we only consider fully spin-polarised Fermions (i.e. all particles are considered to have the same spin). 

Statistical methods were intentionally chosen which would scale to large systems: though we use a dense histogram structure to save the permutation-family binned data, this is not required by any of the underlying algorithms. 
As recognised by DBA it key to correctly predict the behaviour of permutation-families which are so rare they do not get visited in finite time, as we still need to sum over these regions (where we have no data) to produce the final estimate. 

A key practical issue was how to represent the permutation-families. 
Our initial attempt used the $\lambda$ representation from mathematical partition theory (defined as counting number of particles in each cycle, so $\sum_{k=1}^{N} \lambda_k = N$), however ranking this object lexicographically (for indexing the permutation families) was inefficient, and required the building and traversal of a tree structure when building the machine-learning models (to ensure conservation of particle number / probability). 
The very similar $C_k$ vector representation (where each element counts the number of cycles, therefore $\sum_{k=1}^{N} C_k * k = N$) was much more well behaved, as a simple $N \times N$ matrix constructed with dynamic programming permits an $\mathcal{O}(N)$ traversal to rank and unrank, and a similar trivial lookup for masking which values next in the vector are permitted. 

For computational efficiency, our current implementation creates a dense instantiation of the permutation-families. 
This practically limits simulation sizes to $N\approx100$, where the memory required for this object is 3 GB (a \textsc{Float64} and an \textsc{Int64} for each histogram point, so 16 bytes for each $p(N=100)=1.9\times10^8$ bin).

\section{Results}\label{results}

We test the approach on a  homogeneous system (spin-polarised uniform electron gas (UEG)), and a highly inhomogeneous system (spin-polarised electrons in a two-dimensional harmonic trap). 
We ask three questions: 
Does our method agree with the direct Monte Carlo ratio estimator where the sign problem is mild---that is, do our models introduce bias? 
How far does the minimal linear model take us? 
And where must we learn the correlations \textit{between} exchange cycles? 

Rather than attempt to converge new reference data, we run short simulations and compare the convergence efficiency of the na\"ive Monte Carlo Fermionic ratio estimator, with the hierarchy of Machine Learning approaches. 

\subsection{Spin-polarised Uniform Electron Gas}\label{results-ueg}

We implement a three dimensional spin-polarised uniform electron gas (UEG) with the Yakub-Ronchi\cite{Yakub2003} spherically averaged interaction potential and Kelbg short-range corrections\cite{Filinov2003}.

This consists of $N$ electrons in a periodic cubic cell with a neutralising background,

\begin{equation}
    \hat{H}_{\mathrm{UEG}} = -\frac{1}{2}\sum_{i=1}^{N} \nabla_i^2
    + \sum_{i<j} v_{\mathrm{YR}}(r_{ij}).
\end{equation}

Here $v_{\mathrm{YR}}$ is the spherically averaged
Yakub--Ronchi pair potential\cite{Yakub2003} with Kelbg short-range quantum
smoothing\cite{Filinov2003}, following Dornheim\cite{dornheimPathIntegralMonte2019}. 
The system is specified by the density $r_s = (3/4\pi n)^{1/3}$ (in units of the Bohr radius) and the reduced temperature $\Theta = T/T_F$. 
Energies are reported in Rydberg per particle for comparison with DBA\cite{dubois2017overcoming}, whereas internally the code uses Hartree as the natural units of the Hamiltonian.

We use magic-number particle counts $N \in \{7, 19, 33\}$ (closed-shell filling of the three-dimensional Fermi sphere), and sweep the reduced temperature $\theta = T/T_F \in \{1.0, 0.125\}$ and density parameter $r_s \in \{1.0, 10.0\}$.
The number of permutation families grows rapidly with system size: $p(7)=15$, $p(19)=490$, $p(33)=10\,143$.
One caution on our absolute energies: the reference values for $N=33$ are from DBA\cite{dubois2017overcoming}, who use a full Ewald summation, whereas the Yakub-Ronchi potential is an essentially pairwise approximation.
These approaches seem to agree well at low density but start divergong by $r_s=1$, where the periodic corrections matter; comparisons at high density can therefore only test internal consistency rather than absolute accuracy.

Table~\ref{tab:ueg-results} collects the energy estimators from well converged ($100 \times 10^6$ Monte Carlo step) simulations across the four quadrants of the $(r_s, \theta)$ parameter space explored in DBA; Table~\ref{tab:ueg-convergence-n19} follows the $N=19$ system as the budget grows from $1 \times 10^6$ to $300 \times 10^6$ steps (the corresponding $N=7$ and $N=33$ sweeps are in Appendix~\ref{app:ueg-convergence}).
Each energy model (the direct Monte Carlo ratio $E_{\text{MC}}$, the linear model $E_{\text{Lin}}$, and the deep network $E_{\text{Deep}}$) is evaluated against both the MAP-on-DuBois and the LSTM permutation-family probability models.

Where the sign problem is mild, everything agrees.
At low density and warm temperature ($r_s=10$, $\theta=1$) the electrons are well separated, exchange cycles are rare, and the average sign stays large ($\bar{\sigma} \ge 0.4$ even at $N=33$).
The na\"ive ratio estimator is well defined here, and every combination of energy and probability model reproduces the same values. 
Data from this quadrant establishes that the inference machinery introduces no detectable bias where the answer is well defined.

Away from this benign corner, the main finding for the homogeneous gas is that a minimal linear energy model is all that is required.

The linear model (a MAP estimator carrying between $d_{eff} \approx 2$ and $18$ effective degrees of freedom, depending on noise and permutation complexity) and the fully expressive deep network converge to the same energies, within $\sim 0.05 \, \text{Ry}/N$, wherever both are stable.

In a homogeneous system there is no spatial structure to tie the energy of a permutation family to its particular exchange topology, so there are no residual correlations for the network to learn.
Expressiveness (i.e. a more powerful model) is not just unnecessary, but is actively harmful in the quality of the estimator.
At $N=19$, $r_s=1$, $\theta=1$ ($\bar{\sigma} \approx 0.03$), the linear model evaluated on the MAP-on-DuBois probability model converges smoothly ($8.3 \to 8.7 \to 8.6 \to 8.6 \to 8.7 \, \text{Ry}/N$ from $1$M to $300$M steps), whereas the same energy data evaluated on the LSTM probability model swings wildly at intermediate budgets (reaching $+19.9 \, \text{Ry}/N$ at $100$M) and only settles by $300$M. 
Given a poorly sampled permutation histogram, the LSTM fits the sampling noise; the more rigid analytic prior provides regularisation.

The linear model remains serviceable deep into the sign collapse.
At low density and cold temperature ($r_s=10$, $\theta=0.125$) the sign is essentially zero for all $N$, yet $E_{\text{Lin}}$ converges to $-0.198 \, \text{Ry}/N$ at $N=33$, within $5\%$ of the DBA reference of $-0.208 \, \text{Ry}/N$.
At high density and warm temperature ($r_s=1$, $\theta=1$, $N=33$), where the sign has collapsed completely, the linear model stays internally self-consistent at $8.5$--$8.8 \, \text{Ry}/N$ across all computational budgets, while the raw ratio estimator wanders over a range of $10 \, \text{Ry}/N$ and even changes sign.
(The DBA reference here is $17.38 \, \text{Ry}/N$; as noted above, this disagreement hopefully reflects the Ewald vs. Yakub-Ronchi differences of the potential, rather than bugs in our codes.)

Our approach fails in the high-density cold quadrant ($r_s=1$, $\theta=0.125$), which DBA were able to converge.
Here the raw Monte Carlo variance explodes and destroys the regression targets themselves ($R^2 < 5\%$ throughout); the linear model collapses to $E_{\text{Lin}} \approx -1.7 \, \text{Ry}/N$. 
These runs did not use importance sampling, and the missing Ewald corrections is strongest here, which perhaps somewhat changes the average sign making it more difficult to converge.

\begin{table*}[htbp]
    \centering
    \begin{tabular}{c c c r | r | r r r | r r r}
        \hline\hline
        & & & & & \multicolumn{3}{c|}{MAP-on-DuBois prob.\ model} & \multicolumn{3}{c}{LSTM prob.\ model} \\
        \cline{6-11}
        $N$ & $\theta$ & $r_s$ & Steps & $\bar{\sigma}$ & $E_{\text{MC}}/N$ & $E_{\text{Lin}}/N$ & $E_{\text{Deep}}/N$ & $E_{\text{MC}}/N$ & $E_{\text{Lin}}/N$ & $E_{\text{Deep}}/N$ \\
        \hline
        \multicolumn{11}{c}{\emph{Low density ($r_s = 10$), warm ($\theta=1$)}} \\
        \hline
        7  & 1.0 & 10 & $100$M & $+0.82$ & $-0.049$ & $-0.049$ & $-0.049$ & $-0.049$ & $-0.049$ & $-0.049$ \\
        19 & 1.0 & 10 & $100$M & $+0.60$ & $-0.046$ & $-0.046$ & $-0.046$ & $-0.046$ & $-0.046$ & $-0.046$ \\
        33 & 1.0 & 10 & $100$M & $+0.40$ & $-0.046$ & $-0.045$ & $-0.045$ & $-0.045$ & $-0.045$ & $-0.045$ \\
        \hline
        \multicolumn{11}{c}{\emph{Low density ($r_s = 10$), cold ($\theta=0.125$)}} \\
        \hline
        7  & 0.125 & 10 & $100$M & $+0.00$ & $-0.147$ & $-0.164$ & $-0.164$ & $-0.186$ & $-0.186$ & $-0.186$ \\
        19 & 0.125 & 10 & $100$M & $-0.00$ & $-0.188$ & $-0.193$ & $-0.193$ & $-0.192$ & $-0.192$ & $-0.192$ \\
        33 & 0.125 & 10 & $100$M & $-0.00$ & $-0.662$ & $-0.198$ & $-0.205$ & $-0.196$ & $-0.197$ & $-0.197$ \\
        \hline
        \multicolumn{11}{c}{\emph{High density ($r_s = 1$), warm ($\theta=1$)}} \\
        \hline
        7  & 1.0 & 1 & $100$M & $+0.31$ & $+8.42$  & $+8.56$ & $+8.56$  & $+8.44$  & $+8.44$ & $+8.44$ \\
        19 & 1.0 & 1 & $100$M & $+0.03$ & $+8.89$  & $+8.61$ & $+8.62$  & $+19.94$ & $+19.95$ & $+19.95$ \\
        33 & 1.0 & 1 & $100$M & $+0.00$ & $-1.03$  & $+8.72$ & $+6.40$  & $+7.59$  & $+7.90$ & $+7.90$ \\
        \hline
        \multicolumn{11}{c}{\emph{High density ($r_s = 1$), cold ($\theta=0.125$)}} \\
        \hline
        7  & 0.125 & 1 & $100$M & $+0.00$ & $-83.2$ & $-0.76$ & $-0.76$ & $-4.90$ & $-4.90$ & $-4.90$ \\
        19 & 0.125 & 1 & $100$M & $-0.00$ & $-2.19$ & $-1.71$ & $-1.94$ & $-1.70$ & $-1.75$ & $-1.75$ \\
        33 & 0.125 & 1 & $100$M & $+0.00$ & $-2.31$ & $-1.80$ & $+1.01$ & $-1.73$ & $-1.52$ & $-1.52$ \\
        \hline
        \multicolumn{11}{c}{\emph{Reference: DuBois et al.\! \cite{dubois2017overcoming} ($N=33$)}} \\
        \hline
        33 & 1.0   & 1  & \multicolumn{2}{r|}{} & \multicolumn{3}{c|}{$+17.38$} & \multicolumn{3}{c}{} \\
        33 & 1.0   & 10 & \multicolumn{2}{r|}{} & \multicolumn{3}{c|}{$-0.081$} & \multicolumn{3}{c}{} \\
        33 & 0.125 & 1  & \multicolumn{2}{r|}{} & \multicolumn{3}{c|}{$+4.70$}  & \multicolumn{3}{c}{} \\
        33 & 0.125 & 10 & \multicolumn{2}{r|}{} & \multicolumn{3}{c|}{$-0.208$} & \multicolumn{3}{c}{} \\
        \hline\hline
    \end{tabular}

    \caption{Homogeneous UEG: energy estimators in Ry/$N$ at $100$M steps. Each energy head (MC ratio, linear, deep) is evaluated against two probability models: the analytic MAP-on-DuBois prior and the learned LSTM. DuBois reference values~\cite{dubois2017overcoming} use Ewald summation at $N=33$ and are not directly comparable at $r_s=1$.}
    \label{tab:ueg-results}
\end{table*}

A practical benefit of the linear model is in how quickly it converges, quantified where the sign problem is non-trivial in Table~\ref{tab:ueg-convergence-n19}.
For $N=19$ at $r_s=1$, $\theta=1$ ($490$ families), the raw ratio estimator oscillates across a range of $35 \, \text{Ry}/N$ (from $-24.3$ to $+10.9$) as the budget grows, while the linear model varies by only $0.4 \, \text{Ry}/N$; it is within $3\%$ of its converged value from the first million steps, roughly a tenth of the budget the ratio estimator needs to approach consistency.
The same pattern holds for $N=33$ ($10\,143$ families; Appendix~\ref{app:ueg-convergence}), where the linear model settles at $\approx 8.7 \, \text{Ry}/N$ from $1$M steps, insensitive to the oscillations in the measured sign.

For a homogenous system, the permutation-family structure has, in effect, converted the Fermion-sign cancellation problem into a well conditioned regression.

\begin{table}[htbp]
    \centering
    \resizebox{\columnwidth}{!}{
    \begin{tabular}{c c c r | r | r | r | r c | r c}
        \hline\hline
        $N$ & $\theta$ & $r_s$ & Steps & $\bar{\sigma}$ & $\text{KL}_{\text{MAP}}$ & $E_{\text{MC}}/N$ & \multicolumn{2}{c|}{$E_{\text{Lin}}^{\text{MAP}}/N$ ($R^2$)} & \multicolumn{2}{c}{$E_{\text{Deep}}^{\text{LSTM}}/N$ ($R^2$)} \\
        \hline
        \multicolumn{11}{c}{\emph{$N=19$, $r_s=10$, $\theta=1.0$ (Low density, warm)}} \\
        \hline
        19 & 1.0 & 10 & $1$M & $+0.64$ & 0.0005 & $-0.05$ & $-0.05$ & (54\%) & $-0.05$ & (99\%) \\
        19 & 1.0 & 10 & $3$M & $+0.53$ & 0.0065 & $-0.05$ & $-0.05$ & (66\%) & $-0.05$ & (89\%) \\
        19 & 1.0 & 10 & $10$M & $+0.58$ & 0.0012 & $-0.05$ & $-0.05$ & (67\%) & $-0.05$ & (94\%) \\
        19 & 1.0 & 10 & $30$M & $+0.59$ & 0.0025 & $-0.05$ & $-0.05$ & (90\%) & $-0.05$ & (96\%) \\
        19 & 1.0 & 10 & $100$M & $+0.60$ & 0.0025 & $-0.05$ & $-0.05$ & (99\%) & $-0.05$ & (97\%) \\
        19 & 1.0 & 10 & $300$M & $+0.60$ & 0.0021 & $-0.05$ & $-0.05$ & (98\%) & $-0.05$ & (97\%) \\
        \hline
        \multicolumn{11}{c}{\emph{$N=19$, $r_s=10$, $\theta=0.125$ (Low density, cold)}} \\
        \hline
        19 & 0.125 & 10 & $1$M & $+0.02$ & 0.4105 & $-0.21$ & $-0.19$ & (4\%) & $-0.18$ & (72\%) \\
        19 & 0.125 & 10 & $3$M & $-0.00$ & 0.1891 & $-0.39$ & $-0.19$ & (8\%) & $-0.19$ & (33\%) \\
        19 & 0.125 & 10 & $10$M & $+0.02$ & 0.1015 & $-0.18$ & $-0.19$ & (18\%) & $-0.19$ & (40\%) \\
        19 & 0.125 & 10 & $30$M & $-0.00$ & 0.0919 & $-0.19$ & $-0.19$ & (22\%) & $-0.19$ & (36\%) \\
        19 & 0.125 & 10 & $100$M & $-0.00$ & 0.0667 & $-0.19$ & $-0.19$ & (46\%) & $-0.19$ & (72\%) \\
        19 & 0.125 & 10 & $300$M & $-0.00$ & 0.0645 & $-0.18$ & $-0.19$ & (54\%) & $-0.19$ & (91\%) \\
        \hline
        \multicolumn{11}{c}{\emph{$N=19$, $r_s=1$, $\theta=1.0$ (High density, warm)}} \\
        \hline
        19 & 1.0 & 1 & $1$M & $+0.04$ & 0.0277 & $+10.90$ & $+8.29$ & (51\%) & $+10.50$ & (37\%) \\
        19 & 1.0 & 1 & $3$M & $-0.00$ & 0.0183 & $-24.30$ & $+8.71$ & (68\%) & $-34.52$ & (47\%) \\
        19 & 1.0 & 1 & $10$M & $+0.06$ & 0.0135 & $+8.22$ & $+8.69$ & (79\%) & $+9.11$ & (66\%) \\
        19 & 1.0 & 1 & $30$M & $+0.03$ & 0.0103 & $+8.52$ & $+8.64$ & (88\%) & $+15.61$ & (82\%) \\
        19 & 1.0 & 1 & $100$M & $+0.03$ & 0.0086 & $+8.89$ & $+8.61$ & (95\%) & $+19.95$ & (93\%) \\
        19 & 1.0 & 1 & $300$M & $+0.04$ & 0.0084 & $+8.31$ & $+8.65$ & (98\%) & $+8.54$ & (98\%) \\
        \hline
        \multicolumn{11}{c}{\emph{$N=19$, $r_s=1$, $\theta=0.125$ (High density, cold)}} \\
        \hline
        19 & 0.125 & 1 & $1$M & $+0.01$ & 0.1429 & $+4.21$ & $-1.92$ & (3\%) & $+1.61$ & (54\%) \\
        19 & 0.125 & 1 & $3$M & $-0.00$ & 0.0705 & $+2.98$ & $-1.70$ & (1\%) & $-0.64$ & (38\%) \\
        19 & 0.125 & 1 & $10$M & $-0.01$ & 0.0468 & $-0.29$ & $-1.69$ & (2\%) & $-1.20$ & (32\%) \\
        19 & 0.125 & 1 & $30$M & $-0.00$ & 0.0359 & $-2.80$ & $-1.72$ & (12\%) & $-0.87$ & (24\%) \\
        19 & 0.125 & 1 & $100$M & $-0.00$ & 0.0285 & $-2.19$ & $-1.71$ & (26\%) & $-1.75$ & (56\%) \\
        19 & 0.125 & 1 & $300$M & $+0.00$ & 0.0286 & $+2.34$ & $-1.70$ & (36\%) & $-1.77$ & (77\%) \\
        \hline\hline
    \end{tabular}
    }
    \caption{Homogeneous UEG: convergence with Monte Carlo budget for $N=19$ ($p(19)=490$ families) across the four $(r_s, \theta)$ quadrants. Energies in Ry/$N$. $\text{KL}_{\text{MAP}}$ is the divergence of the MAP-on-DuBois prior from the empirical permutation-family frequencies; $R^2$ is the coefficient of determination of the energy fit. The $N=7$ and $N=33$ equivalents are in Appendix~\ref{app:ueg-convergence}.}
    \label{tab:ueg-convergence-n19}
\end{table}

\subsection{Electrons in a harmonic trap (quantum dot)}\label{results-harmonictrap}

Following Dornheim\cite{Dornheim2019-UEG-2DTRAP} we consider $N$ spin-polarised electrons in a two-dimensional harmonic potential, a simple model of a quantum dot. 
Our Hamiltonian is 

\begin{equation}
    \hat{H}_{\mathrm{trap}} = \sum_{i=1}^{N}
    \left( -\frac{1}{2}\nabla_i^2 + \frac{1}{2} r_i^2 \right)
    + \sum_{i<j} \frac{\lambda}{r_{ij}}.
\end{equation}

Energy is measured in units of the oscillator energy $\hbar\omega$, which thereby defines length as $\ell_0 = \sqrt{\hbar/m\omega}$. 
Our system is specified by the inverse temperature $\beta$ in units of $1/\hbar\omega$, and 
the dimensionless coupling $\lambda = e^2 / (\ell_0\, \hbar\omega)$. 
This coupling controls the crossover from a fully delocalised dot of noninteracting electrons ($\lambda \to 0$) to a Wigner molecule ($\lambda \gg 1$).

Again we use magic-numbers for particle filling in a 2D harmonic trap, $N \in \{6, 10, 20\}$, consider inverse temperatures $\beta \in \{0.3, 1.0\}$ (Table~\ref{tab:harmonic-trap-results}), and vary the Coulomb coupling $\lambda$. 

The harmonic trap is implicitly inhomogeneous due to spatial structure, a long exchange cycle must thread through the confining potential, and the energy it accrues depends on where it sits; the energies of different cycles in the same configuration are therefore correlated. 
This is the statistical dependence between cycle lengths that Dornheim et al.\cite{Dornheim2019permutation} identified, and it is what the separable linear models cannot represent. 
We should therefore expect the linear models to break down here, and the correlation-learning LSTM models to become necessary. 
After some tuning to get the models to successfully train, this is indeed what we observe. 

\begin{figure}
    \centering
    \includegraphics[width=1.0\linewidth]{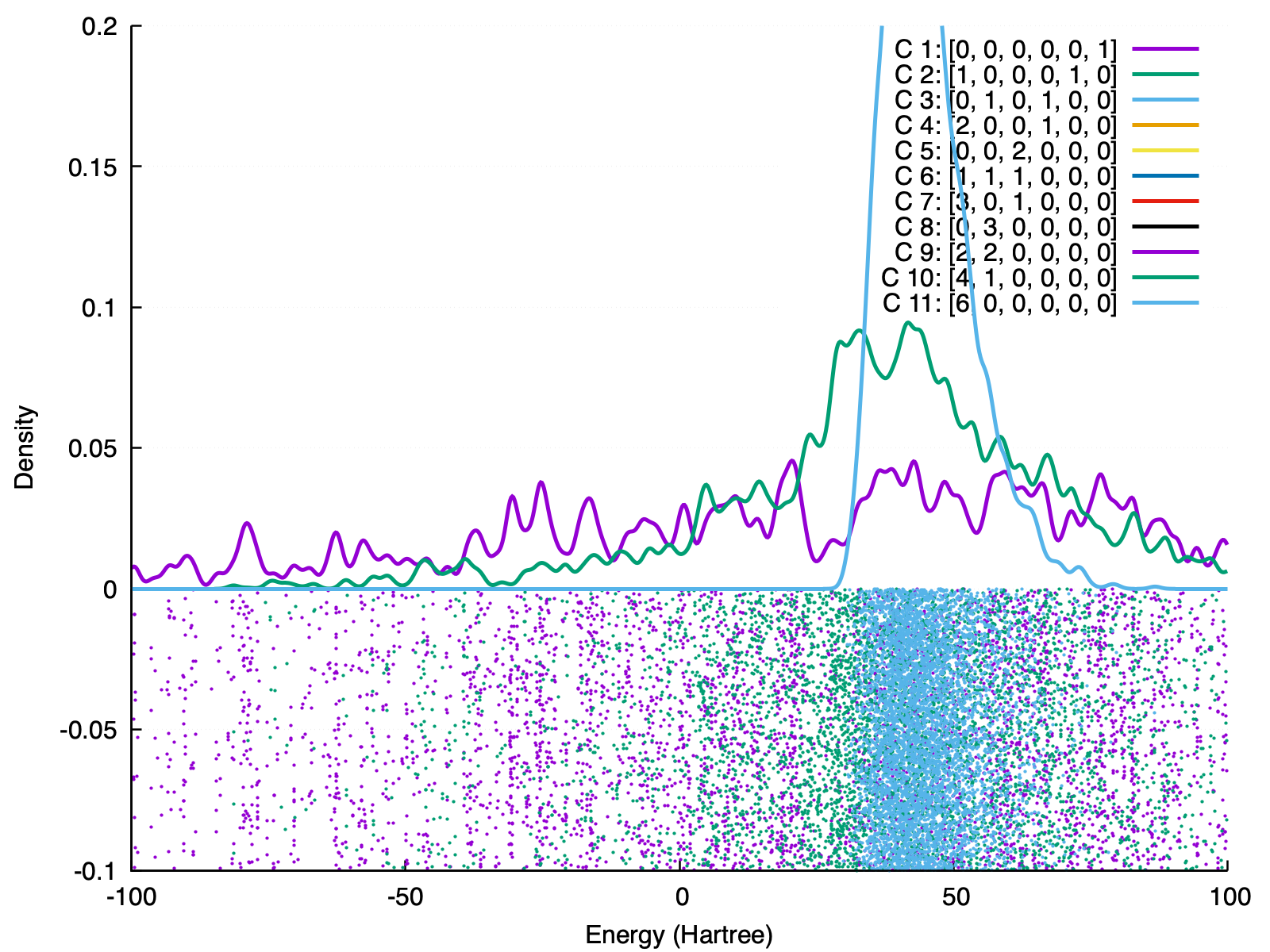}
    \includegraphics[width=1.0\linewidth]{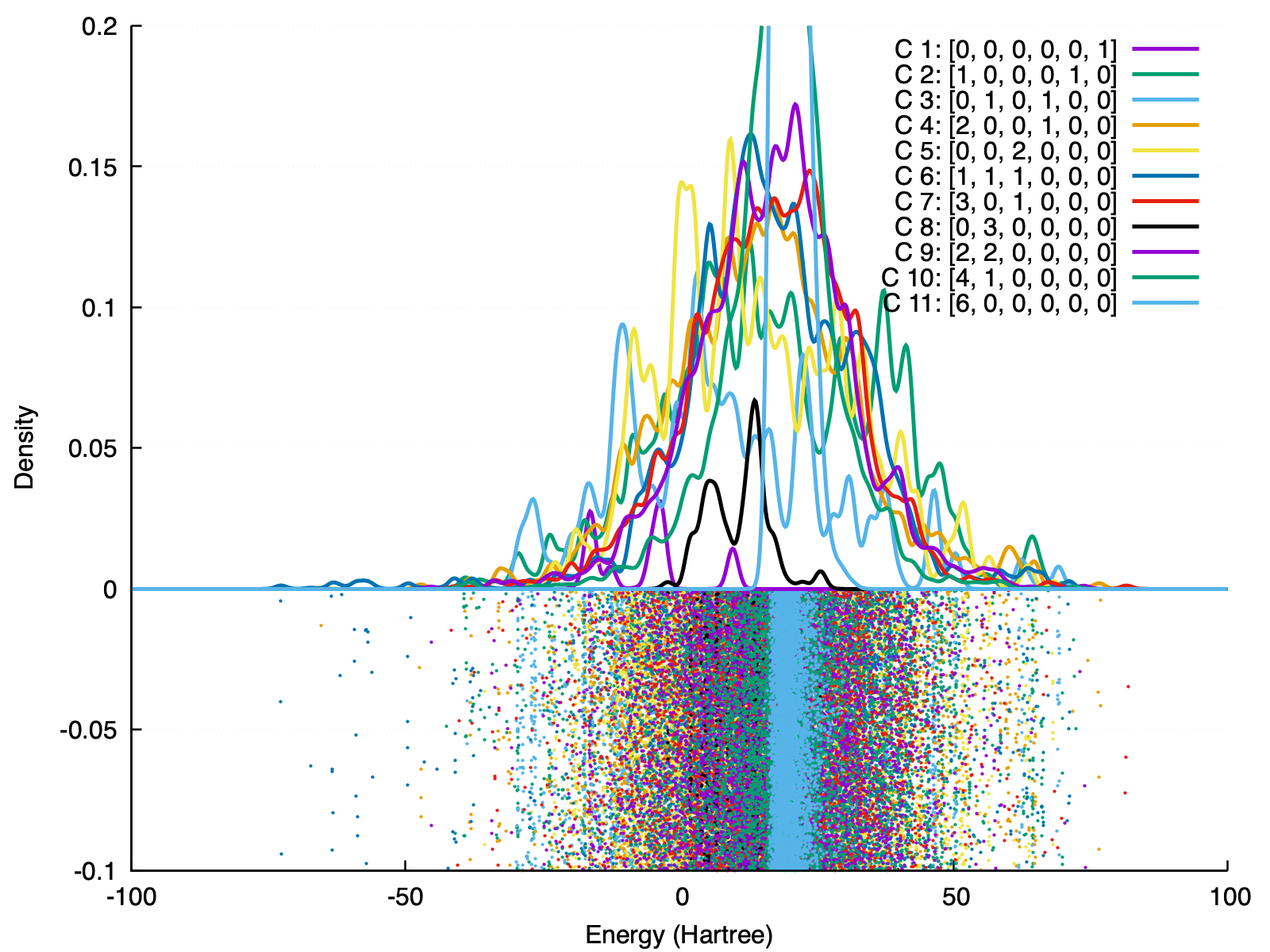}

    \caption{
    Up to 1000 energy samples from each permutation-family in an $N=6$ 2D harmonic trap at higher temperature $\beta=0.3$ (top) and lower temperature $\beta=1.0$ (bottom). 
    The 11 permutation-families are explicitly labelled with their permutation vector. 
    It is very noticeable that in this system the energies are not symmetrically distributed around a mean nor Gaussian. 
    Particularly, the trivial $6 \times 1$-mer permutation cycle ($C: \, 11$) is highly skewed, and some degree of banding structure is visible in the larger permutations.
    }
    \label{fig:trap-reservoir}
\end{figure}

The contrast with the homogeneous gas is visible directly in the raw energy samples
(Figure~\ref{fig:trap-reservoir}): the within-family energy distributions in
the trap are skewed and multi-modal, which a separable (linear) model would not be able to represent. 

For a small, warm system ($N=6$, $\beta=0.3$) the average sign remains high and, as in the benign corner of the UEG, all estimators simply verify one another: the de-biased Monte Carlo, the rigid linear model, and the LSTM models agree to well within $1 \, \hbar\omega$ of $46.5 \, \hbar\omega$. 
Cooling to $\beta=1.0$ drops the sign to $\bar{\sigma} \approx 0.05$ and the models begin to separate: the rigid linear model drifts, while the LSTM-driven estimates remain tight ($21.8$--$21.9 \, \hbar\omega$) and consistent with the Monte Carlo baseline. 

At $N=10$, $\beta=0.3$ the raw simulations visit only 13 of the 42 permutation families, but the sign remains manageable and all models concur around $85$--$86 \, \hbar\omega$. 
Cooling to $\beta=1.0$ collapses the sign entirely ($\bar{\sigma} \approx -0.005$ in the biased run), and the ratio estimator becomes worthless; the LSTM probability model nevertheless relaxes onto sensible sector probabilities, and the deep energy model extracts stable observables from data the ratio estimator cannot use. 

Our most severe test is $N=20$ at $\beta=0.3$, with 627 permutation families. 
The unbiased Monte Carlo fails outright ($\bar{\sigma} \approx -0.012$, wild estimates of $\approx 201 \, \hbar\omega$). 
Activating the permutation-family importance sampling ($\alpha=0.7$, driven by the jackknife standard errors of Section~\ref{importance-sampling}) pushes the simulation into a regime with a small but resolvable positive sign ($\bar{\sigma} \approx 0.04$). 
Retraining the models on this targeted data, the LSTMs interpolate across the unvisited families and yield $182.2 \, \hbar\omega$ (shallow) and $183.7 \, \hbar\omega$ (deep). 
A regime inaccessible to raw path-integral Monte Carlo is thus recovered by steering the sampling with the learned probability model, with agreement between the independent model heads as the internal check on the result. 

\begin{table*}[htbp]
    \centering
    \begin{tabular}{c c c | r | r c | r l r l r l}
        \hline\hline
        & & & \multicolumn{1}{c|}{Raw Data} & \multicolumn{2}{c|}{IS Diagnostics} & \multicolumn{6}{c}{IS $E/\hbar\omega$ Estimators \& Quality ($R^2$)} \\
        \cline{4-6} \cline{7-12}
        $N$ & $\beta$ & $\bar{\sigma}$ & \multicolumn{1}{c|}{Raw MC} & $\text{KL}$ & $d_{eff}$ & \multicolumn{2}{c}{\text{Linear Model}} & \multicolumn{2}{c}{\text{Shallow LSTM}} & \multicolumn{2}{c}{\text{Deep LSTM}} \\
        \hline
        6  & 0.3 &  0.662 &  46.49 & 0.000 & 2.0 &  46.60 & (99.5\%) & 46.58 & (98.7\%) & 46.59 & (98.7\%) \\
        6  & 1.0 &  0.049 &  22.38 & 0.010 & 5.8 &  21.79 & (99.9\%) & 21.84 & (99.9\%) & 21.80 & (99.9\%) \\
        10 & 0.3 &  0.246 &  86.34 & 0.003 & 2.0 &  87.32 & (99.1\%) & 86.34 & (97.8\%) & 86.35 & (97.7\%) \\
        10 & 1.0 &  0.016 &  36.53 & 0.052 & 9.1 &  36.21 & (98.6\%) & 37.36 & (97.9\%) & 36.59 & (97.9\%) \\
        20 & 0.3 & -0.012 & 201.49 & 0.036 & 2.0 & 177.36 & (96.9\%) & 182.18 & (91.2\%) & 183.66 & (92.1\%) \\
        \hline\hline
    \end{tabular}
    \caption{2D harmonic trap: summary estimators for $N$ electrons at inverse temperature $\beta$. $\bar{\sigma}$ and the `Raw MC' energy are from the unbiased Monte Carlo run; the three energy heads (linear, shallow LSTM, deep LSTM, with $R^2$) are fitted to the importance-sampled (IS) data. KL is the divergence of the analytic DuBois prior from the data, and $d_{eff}$ the effective degrees of freedom of the ridge regression.}
    \label{tab:harmonic-trap-results}
\end{table*}

The inference framework we developed also reports on its own health. 
Table~\ref{tab:harmonic-trap-results} includes two internal diagnostics: the Kullback-Leibler divergence between the analytic physics prior and the empirically observed permutation frequencies, and the effective degrees of freedom $d_{eff}$ retained by the ridge regression. 
In the warm systems the divergence is minimal ($\approx 0.001$): the analytic prior already describes the data. 
In the regimes of severe sign collapse it jumps by more than an order of magnitude, warning that the simple separable models no longer describe the permutation landscape and that the flexible models are carrying the fit. 

\subsection{Coulomb coupling sweep: the Wigner molecule crossover}

A more demanding test than any single state point is to track an observable across a continuous crossover. 
By varying the Coulomb coupling, we can push the system from a localised Wigner molecule (with electrons crystallised out) to a fully delocalised quantum dot. 
We sweep the Coulomb coupling strength $\lambda$ for $N=6$ at $\beta=1.0$ using $10^9$ Monte Carlo steps (Table~\ref{tab:lambda-sweep-results}).
The sweep is also an alternate probe of the sign problem itself, since the coupling directly modulates the prevalence of exchange.

At strong coupling ($\lambda \ge 3$) the Coulomb repulsion separates the electrons in space, exchange cycles are exponentially suppressed, the sign stays near unity, and all estimators agree trivially.
As the coupling weakens through $\lambda \in [0.3, 1.0]$ the particles delocalise and overlap at the trap centre, triggering multi-particle ring exchanges; the sign collapses from $\bar{\sigma} = 0.14$ to $0.014$.
Across this crossover the energy models remain mutually consistent and physically sensible: at $\lambda=0.5$ the linear and deep models both give $22.6 \, \hbar\omega$, and at $\lambda=0.3$ the estimates remain above the non-interacting bound of $E_0^{\text{NI}} = 14 \, \hbar\omega$.

Below $\lambda \approx 0.1$ the near-complete sign collapse ($\bar{\sigma} < 0.01$) corrupts the energy targets themselves.
At $\lambda=0.1$ the linear model, anchored by the Feynman exchange-penalty prior, still returns plausible values while the raw ratio estimator is wild; but by $\lambda = 0.01$ the models produce $E \approx 56 \, \hbar\omega$ against a Bosonic Jackknife bound of $9.5 \, \hbar\omega$. 
The LSTM probability model seems to have become unphysical. 

\begin{table}[htbp]
    \centering
    \begin{tabular}{cr|crr|rr}
        \hline\hline
        $\lambda$ & $\bar{\sigma}$ & $\text{Bos.\!} E \pm \delta E_J$ & $\text{KL}_{\text{MAP}}$ & $E_{\text{MC}}$ & $E_{\text{Lin}}$ & $E_{\text{Deep}}$  \\
        \hline
        10.0  &  0.990 & $74.6 \pm 0.1$ & 0.000 & 74.58 & 74.58 & 74.58 \\
        3.00  &  0.667 & $39.7 \pm 0.3$ & 0.002 & 40.17 & 40.18 & 40.18 \\
        1.00  &  0.143 & $23.9 \pm 0.3$ & 0.009 & 26.66 & 26.61 & 26.60 \\
        0.50  &  0.040 & $17.9 \pm 0.3$ & 0.014 & 23.35 & 22.62 & 22.61 \\
        0.30  &  0.014 & $14.9 \pm 0.2$ & 0.019 & 22.05 & 22.54 & 22.50 \\
        0.10  &  0.008 & $11.3 \pm 0.2$ & 0.023 & 20.01 & 17.73 & 17.70 \\
        0.05  &  0.001 & $10.3 \pm 0.2$ & 0.024 & 47.02 & 35.52 & 35.57 \\
        0.01  &  0.001 & $ 9.5 \pm 0.2$ & 0.023 & 74.40 & 55.99 & 55.99 \\
        0.00  &  0.003 & $ 9.2 \pm 0.1$ & 0.027 & 20.84 & $-$2.75 & $-$2.33 \\
        \hline\hline
    \end{tabular}
    \caption{2D harmonic trap: Coulomb coupling ($\lambda$) sweep for $N=6$ at $\beta=1.0$ ($10^9$ MC steps, unbiased). $\text{Bos.\!} E$ is the sign-free Bosonic Jackknife bound; $E_{\text{Lin}}$ and $E_{\text{Deep}}$ are evaluated against the LSTM probability model. Estimates below the non-interacting ground state $E_0^{\text{NI}} = 14 \, \hbar\omega$ are unphysical.}
    \label{tab:lambda-sweep-results}
\end{table}

\subsection{Scaling towards the exponential wall: temperature and particle number}

How much do these methods practically improve the sign problem in an inhomogenous 2D trap?  
We map performance with inverse temperature $\beta$ and particle number $N$, using extremely moderate simulations of $10\times 10^6$ Monte Carlo steps (Tables~\ref{tab:beta-scaling-results} and~\ref{tab:n-scaling-results}).
All energy models here are evaluated against the LSTM (MAP-on-DuBois prior) probability model, the most expressive in our hierarchy.
The \textit{unbiased} rows use the initial unbiased Monte Carlo run; the importance-sampled (\emph{IS}) rows refits the energy heads on the jackknife-guided biased data, but reuses the \emph{same} frozen LSTM probability model from the unbiased run. 
The non-interacting spin-polarised ground states ($E_0^{\text{NI}} = 14 \, \hbar\omega$ for $N=6$, $30 \, \hbar\omega$ for $N=10$) provides a lower bound, as we would generally expect the repulsive $\lambda=1$ Coulomb interaction to raise the energy.

Three regimes emerge.
Again, where the sign is healthy ($\bar{\sigma} > 0.1$), all estimators agree (i.e. our models no not introduce a general bias).
In an intermediate $0.01 < \bar{\sigma} < 0.1$ regime, the na\"ive Fermionic ratio estimator oscillates considerably, but the model-based estimators---especially with importance sampling---remain consistent. 
At $\beta=1.0$ the importance-sampled deep model improves $R^2 > 99\%$, whereas the fit to unbiased Monte Carlo manages only $95.7\%$.

Generally it seems we have extended the useful operating range of the simulation by roughly a factor of two in the sign collapse, at fixed budget.

Once $\bar{\sigma} < 0.01$, all estimators fail for the moderate budget: at $\beta \ge 5$ the models report $E \approx 13 \, \hbar\omega$, below the non-interacting bound and therefore unphysical.
Rather disturbingly, in this regime the $R^2$ of the fit looks superficially healthy. 
We think this may be due to the the interaction of the sign collapse with the empirical calculation of variance in our energy estimators. 
We used a Huber loss (which becomes linear beyond $1$ Ha) to reject outliers, and this might then be leading the model to unphysically overfit the low-variance statistical noise of short cycles, yielding a superficially good $R^2$ fit, while missing the Fermionic correlation energy. 
Due to lack of time, investigating and correcting this is left for future work. 

Scaling with particle number (Table~\ref{tab:n-scaling-results}) shows that the model can help with a factor or two increase in the sign problem, but not further. 
By $N=10$ ($\bar{\sigma} = 0.02$), the machine-learnt models can no longer produce sensible fits.

\begin{table*}[htbp]
    \centering
    \begin{tabular}{ccr | cc | r | r | rl rl rl}
        \hline\hline
        $N$ & $\beta$ & \textbf{Block} & $\bar{\sigma}$ & $\text{Bosonic } E \pm \delta E_J$ & ${\text{KL}}_{\text{MAP}}$ & $E_{\text{MC}}$ & \multicolumn{2}{c}{\textbf{Linear} ($R^2$)} & \multicolumn{2}{c}{\textbf{Shallow} ($R^2$)} & \multicolumn{2}{c}{\textbf{Deep} ($R^2$)} \\
        \hline
         6 & 0.3 & Unb. & 0.663 & $44.3 \pm 4.7$ & 0.0063 & 46.16 & 46.32 & (99.3\%) & 46.30 & (99.8\%) & 46.30 & (99.7\%) \\
        & & \emph{IS} & - & $40.1 \pm 3.6$ & N/A & 49.41 & 47.30 & (100.0\%) & 47.30 & (100.0\%) & 47.30 & (100.0\%) \\
         6 & 0.6 & Unb. & 0.197 & $25.0 \pm 3.3$ & 0.0203 & 30.09 & 30.45 & (96.1\%) & 29.80 & (98.3\%) & 29.76 & (98.1\%) \\
        & & \emph{IS} & - & $21.7 \pm 3.7$ & N/A & 28.01 & 30.61 & (98.2\%) & 31.05 & (99.5\%) & 31.16 & (99.5\%) \\
         6 & 0.8 & Unb. & 0.066 & $20.5 \pm 2.9$ & 0.0196 & 23.97 & 27.05 & (95.3\%) & 26.02 & (98.4\%) & 27.00 & (97.7\%) \\
        & & \emph{IS} & - & $18.2 \pm 3.4$ & N/A & 23.54 & 24.67 & (93.4\%) & 26.50 & (98.3\%) & 26.19 & (98.7\%) \\
         6 & 1.0 & Unb. & 0.085 & $18.2 \pm 1.9$ & 0.0145 & 19.04 & 20.11 & (93.8\%) & 19.99 & (95.7\%) & 20.11 & (95.7\%) \\
        & & \emph{IS} & - & $16.0 \pm 2.1$ & N/A & 13.70 & 20.45 & (98.0\%) & 20.83 & (99.7\%) & 20.85 & (99.7\%) \\
         6 & 1.3 & Unb. & 0.046 & $15.7 \pm 2.2$ & 0.0311 & 19.54 & 17.15 & (96.0\%) & 18.88 & (98.9\%) & 18.58 & (99.0\%) \\
        & & \emph{IS} & - & $14.7 \pm 1.9$ & N/A & 17.39 & 17.06 & (98.2\%) & 16.20 & (99.0\%) & 16.51 & (99.2\%) \\
         6 & 1.5 & Unb. & -0.039 & $14.9 \pm 1.7$ & 0.0266 & 18.52 & 14.91 & (91.1\%) & 15.14 & (97.7\%) & 14.86 & (97.5\%) \\
        & & \emph{IS} & - & $14.4 \pm 1.7$ & N/A & 15.96 & 15.59 & (98.5\%) & 15.60 & (99.6\%) & 15.73 & (99.7\%) \\
         6 & 2.0 & Unb. & 0.023 & $13.8 \pm 1.7$ & 0.0272 & 21.12 & 15.66 & (90.6\%) & 17.69 & (98.2\%) & 15.64 & (97.5\%) \\
        & & \emph{IS} & - & $13.6 \pm 1.4$ & N/A & 14.72 & 15.09 & (99.0\%) & 15.19 & (99.7\%) & 15.18 & (99.7\%) \\
         6 & 3.0 & Unb. & -0.060 & $13.3 \pm 1.1$ & 0.0112 & 13.84 & 14.08 & (94.5\%) & 13.98 & (98.0\%) & 14.11 & (97.2\%) \\
        & & \emph{IS} & - & $13.3 \pm 1.1$ & N/A & 5.89 & 12.49 & (96.9\%) & 12.31 & (99.2\%) & 12.45 & (98.9\%) \\
         6 & 5.0 & Unb. & 0.025 & $13.0 \pm 0.8$ & 0.0050 & 9.73 & 11.90 & (61.7\%) & 11.50 & (77.9\%) & 11.91 & (70.5\%) \\
        & & \emph{IS} & - & $13.1 \pm 0.9$ & N/A & 12.43 & 13.70 & (94.3\%) & 13.72 & (96.3\%) & 13.65 & (96.4\%) \\
         6 & 10.0 & Unb. & 0.018 & $12.9 \pm 0.7$ & 0.0035 & 11.81 & 12.93 & (49.6\%) & 12.91 & (44.3\%) & 12.93 & (43.7\%) \\
        & & \emph{IS} & - & $12.9 \pm 0.9$ & N/A & 11.05 & 13.07 & (99.1\%) & 13.01 & (99.3\%) & 13.06 & (99.2\%) \\
        \hline\hline
    \end{tabular}
    \caption{2D harmonic trap: temperature ($\beta$) scaling of $E/\hbar\omega$ for $N=6$, $\lambda=1$ ($10\times 10^6$ steps), for unbiased and importance-sampled (IS) blocks, evaluated against the LSTM (MAP-on-DuBois) probability model. Entries below the non-interacting bound $E_0^{\text{NI}} = 14 \, \hbar\omega$ are unphysical.}
    \label{tab:beta-scaling-results}
\end{table*}

\begin{table*}[htbp]
    \centering

    \begin{tabular}{ccr | cc | r | r | rl rl rl}
        \hline\hline
        $N$ & $\beta$ & \textbf{Block} & $\bar{\sigma}$ & $\text{Bosonic } E \pm \delta E_J$ & ${\text{KL}}_{\text{MAP}}$ & $E_{\text{MC}}$ & \multicolumn{2}{c}{\textbf{Linear} ($R^2$)} & \multicolumn{2}{c}{\textbf{Shallow} ($R^2$)} & \multicolumn{2}{c}{\textbf{Deep} ($R^2$)} \\
        \hline
         3 & 1.0 & Unb. & 0.406 & $7.4 \pm 0.7$ & 0.0017 & 9.10 & 9.03 & (99.7\%) & 9.04 & (99.8\%) & 9.04 & (99.8\%) \\
        & & \emph{IS} & - & $6.5 \pm 0.6$ & N/A & 8.48 & 8.86 & (100.0\%) & 8.86 & (100.0\%) & 8.86 & (100.0\%) \\
         4 & 1.0 & Unb. & 0.259 & $10.7 \pm 1.2$ & 0.0155 & 11.95 & 12.38 & (94.3\%) & 12.26 & (97.3\%) & 12.32 & (96.8\%) \\
        & & \emph{IS} & - & $8.7 \pm 1.5$ & N/A & 12.33 & 11.48 & (99.9\%) & 11.67 & (100.0\%) & 11.65 & (100.0\%) \\
               6 & 1.0 & Unb. & -0.028 & $17.8 \pm 2.1$ & 0.0161 & 19.49 & 15.40 & (90.3\%) & 15.56 & (97.4\%) & 14.94 & (97.4\%) \\
        & & \emph{IS} & - & $16.2 \pm 2.9$ & N/A & 20.79 & 15.20 & (99.5\%) & 15.10 & (99.8\%) & 14.94 & (99.8\%) \\
         8 & 1.0 & Unb. & -0.029 & $26.4 \pm 4.5$ & 0.0342 & 22.52 & 20.98 & (87.4\%) & 22.00 & (97.0\%) & 22.18 & (97.0\%) \\
        & & \emph{IS} & - & $24.2 \pm 4.0$ & N/A & 31.00 & 20.55 & (96.9\%) & 22.87 & (98.9\%) & 23.73 & (99.2\%) \\
         10 & 1.0 & Unb. & 0.022 & $36.2 \pm 7.0$ & 0.0746 & 79.37 & 37.04 & (46.6\%) & 80.23 & (91.1\%) & 82.83 & (92.7\%) \\
        & & \emph{IS} & - & $34.3 \pm 7.4$ & N/A & 35.89 & 36.31 & (90.1\%) & 38.31 & (97.8\%) & 43.13 & (99.7\%) \\
         12 & 1.0 & Unb. & -0.014 & $47.6 \pm 10.1$ & 0.1320 & 74.31 & 45.63 & (56.8\%) & 74.30 & (72.0\%) & 72.05 & (75.9\%) \\
        & & \emph{IS} & - & $46.1 \pm 10.9$ & N/A & 36.69 & 44.70 & (76.7\%) & 38.80 & (89.0\%) & 19.09 & (97.2\%) \\
        \hline\hline
    \end{tabular}
    \caption{2D harmonic trap: particle-number ($N$) scaling of $E/\hbar\omega$ at $\beta=1.0$ ($10\times 10^6$ steps). The non-interacting ground states are $E_0^{\text{NI}} = 5, 8, 14, 22, 30, 40 \, \hbar\omega$ for $N=3$--$12$; estimates below these bounds signal unresolved noise.}
    \label{tab:n-scaling-results}
\end{table*}

\subsection{Convergence with computational budget}

Finally, we repeat the computational budget study in the trap, running a difficult but not impossible simulation $N=6$ at $\beta=1.0$ ($\lambda=1$) from $1\times 10^6$ to $300\times 10^6$ steps (Table~\ref{tab:convergence}).

The pattern observed in the electron gas repeats: the raw ratio estimator swings between $25$ and $79 \, \hbar\omega$ (the measured sign itself fluctuates between $-0.05$ and $+0.07$ across independent runs, so any single ratio estimate is unreliable), while the linear model converges smoothly to $\approx 25 \, \hbar\omega$ at roughly a tenth of the budget, returning a physically reasonable value even at $1$M steps.

In the trap, importance sampling of the permutation-family cycles is clearly important, as this stabilises the fit quality from the smallest budgets ($R^2 > 97\%$ even at $1$M); by $300$M all importance-sampled estimators agree at $E \approx 25 \, \hbar\omega$.
The LSTM energy heads again overshoot at intermediate budgets ($52$--$54 \, \hbar\omega$ at $3$M) before settling.

\begin{table*}[htbp]
    \centering
    \begin{tabular}{r | cc | r | rl rl rl}
        \hline\hline
        Steps & $\bar{\sigma}$ & $\text{Bos.\!} E \pm \delta E_J$ & $E_{\text{MC}}$ & \multicolumn{2}{c}{\textbf{Linear} ($R^2$)} & \multicolumn{2}{c}{\textbf{Shallow} ($R^2$)} & \multicolumn{2}{c}{\textbf{Deep} ($R^2$)} \\
        \hline
        \multicolumn{10}{c}{\emph{Unbiased}} \\
        \hline
        $1\times 10^6$   & $-0.048$ & $17.6 \pm 6.0$ & 40.47 & 26.43 & (71.3\%) & 36.68 & (93.0\%) & 39.08 & (93.8\%) \\
        $3\times 10^6$   & $0.008$  & $17.5 \pm 3.6$ & 79.11 & 24.83 & (84.2\%) & 59.59 & (97.1\%) & 66.94 & (97.8\%) \\
        $10\times 10^6$  & $0.072$  & $18.2 \pm 2.4$ & 26.24 & 25.43 & (94.2\%) & 26.00 & (98.7\%) & 25.48 & (98.5\%) \\
        $30\times 10^6$  & $0.026$  & $17.9 \pm 1.4$ & 28.21 & 26.26 & (96.8\%) & 27.27 & (98.6\%) & 26.35 & (98.3\%) \\
        $100\times 10^6$ & $0.010$  & $17.8 \pm 0.8$ & 35.22 & 35.29 & (99.0\%) & 35.80 & (99.7\%) & 35.33 & (99.7\%) \\
        $300\times 10^6$ & $0.024$  & $17.9 \pm 0.4$ & 24.84 & 25.63 & (99.9\%) & 25.41 & (99.9\%) & 25.58 & (99.9\%) \\
        \hline
        \multicolumn{10}{c}{\emph{Importance-Sampled}} \\
        \hline
        $1\times 10^6$   & - & $17.5 \pm 7.1$ & 19.32 & 24.10 & (97.5\%) & 24.11 & (99.4\%) & 22.26 & (99.4\%) \\
        $3\times 10^6$   & - & $16.3 \pm 4.1$ & 16.64 & 34.03 & (97.2\%) & 52.19 & (99.8\%) & 53.61 & (99.8\%) \\
        $10\times 10^6$  & - & $15.5 \pm 2.7$ & 23.31 & 24.21 & (97.1\%) & 26.91 & (99.9\%) & 27.69 & (99.9\%) \\
        $30\times 10^6$  & - & $15.8 \pm 1.3$ & 17.87 & 25.85 & (99.6\%) & 24.71 & (99.9\%) & 25.12 & (99.9\%) \\
        $100\times 10^6$ & - & --- & 23.58 & 33.70 & (99.9\%) & 32.62 & (100.0\%) & 33.74 & (100.0\%) \\
        $300\times 10^6$ & - & --- & 22.03 & 25.25 & (100.0\%) & 24.91 & (100.0\%) & 25.09 & (100.0\%) \\
        \hline\hline
    \end{tabular}
    \caption{2D harmonic trap: convergence of $E/\hbar\omega$ with Monte Carlo budget for $N=6$, $\beta=1.0$, $\lambda=1.0$. Each row is an independent simulation at the stated step count. The non-interacting bound is $E_0^{\text{NI}} = 14 \, \hbar\omega$.}
    \label{tab:convergence}
\end{table*}

\section{Discussion}\label{discussion}

We have implemented a hierarchy of statistically motivated methods to project from Bosonic path integral Monte Carlo to Fermionic observables. 
Our key contribution is showing how physically-motivated minimal parameter models (often essentially mean-field or ideal-gas constructions) can be transparently and productively re-used as an inductive prior in a statistical model that then only has to learn the correlations. 

In noise-dominated regimes the rigid linear model was consistently the most robust estimator. 
Unless the physics demands learned correlations---as it does in inhomogeneous systems---we recommend it as the primary production estimator, with the flexible LSTM models retained as a cross-check. 

There seems to be promise in the permutation-family approach to ameliorate the Fermion sign problem, as all of the models can be fitted in a sign-free region, and refined by importance-sampling and fine-tuning until the required finesse is developed for the extreme cancellation required to head towards the challenging regions of small average sign $\bar\sigma\approx 0$. 

More generally, this work was inspired by the point of view of applying probabilistic numerics to quantum Monte Carlo: by recasting the calculation of quantum observables as an inference problem, we can drastically improve the information efficiency (and therefore signal to noise ratio) of our estimators. 

\section{Acknowledgement}

I am grateful for fruitful discussions with Gabriele Spada and Matthew Foulkes. 
This work was developed for presentation at the April 2026 \textsc{The Sign Problem of Fermions} workshop at ECT*, Villazzano (Trento), Italy; 
I am grateful to the organisers for the motivating invitation.  
J.M.F. 
is supported by a Royal Society University Research Fellowship
(URF-R1-191292). 
I gratefully acknowledge the use of the Imperial College Research Computing Service\cite{HPC}.

\section{Data availability}
The \textsc{Halcyon.jl} code implementing the permutation-family PIMC method
is openly available\cite{Halcyon.jl}, commit d31d95e 28th April 2026. 
The simulation outputs and analysis
scripts underlying Tables~\ref{tab:ueg-results}--\ref{tab:convergence} are
archived at \nb{Zenodo DOI..}.

\appendix

\section{UEG convergence with budget: $N=7$ and $N=33$}\label{app:ueg-convergence}

Tables~\ref{tab:ueg-convergence-n7} and~\ref{tab:ueg-convergence-n33} record the convergence of the UEG estimators with Monte Carlo budget for $N=7$ and $N=33$, complementing the $N=19$ data of Table~\ref{tab:ueg-convergence-n19} in the main text.

\begin{table*}[htbp]
    \centering
    \begin{tabular}{c c c r | r | r | r | r c | r c}
        \hline\hline
        $N$ & $\theta$ & $r_s$ & Steps & $\bar{\sigma}$ & $\text{KL}_{\text{MAP}}$ & $E_{\text{MC}}/N$ & \multicolumn{2}{c|}{$E_{\text{Lin}}^{\text{MAP}}/N$ ($R^2$)} & \multicolumn{2}{c}{$E_{\text{Deep}}^{\text{LSTM}}/N$ ($R^2$)} \\
        \hline
        \multicolumn{11}{c}{\emph{$N=7$, $r_s=10$, $\theta=1.0$ (Low density, warm)}} \\
        \hline
        7 & 1.0 & 10 & $1$M & $+0.85$ & 0.0050 & $-0.05$ & $-0.05$ & (14\%) & $-0.05$ & (99\%) \\
        7 & 1.0 & 10 & $3$M & $+0.80$ & 0.0137 & $-0.05$ & $-0.05$ & (61\%) & $-0.05$ & (96\%) \\
        7 & 1.0 & 10 & $10$M & $+0.82$ & 0.0051 & $-0.05$ & $-0.05$ & (91\%) & $-0.05$ & (99\%) \\
        7 & 1.0 & 10 & $30$M & $+0.82$ & 0.0050 & $-0.05$ & $-0.05$ & (98\%) & $-0.05$ & (100\%) \\
        7 & 1.0 & 10 & $100$M & $+0.82$ & 0.0063 & $-0.05$ & $-0.05$ & (99\%) & $-0.05$ & (98\%) \\
        7 & 1.0 & 10 & $300$M & $+0.82$ & 0.0060 & $-0.05$ & $-0.05$ & (99\%) & $-0.05$ & (100\%) \\
        \hline
        \multicolumn{11}{c}{\emph{$N=7$, $r_s=10$, $\theta=0.125$ (Low density, cold)}} \\
        \hline
        7 & 0.125 & 10 & $1$M & $+0.00$ & 0.0075 & $+0.08$ & $-0.20$ & (29\%) & $-0.17$ & (60\%) \\
        7 & 0.125 & 10 & $3$M & $+0.02$ & 0.0011 & $-0.17$ & $-0.23$ & (44\%) & $-0.17$ & (74\%) \\
        7 & 0.125 & 10 & $10$M & $+0.01$ & 0.0006 & $-0.20$ & $-0.16$ & (14\%) & $-0.17$ & (79\%) \\
        7 & 0.125 & 10 & $30$M & $+0.01$ & 0.0003 & $-0.16$ & $-0.17$ & (33\%) & $-0.17$ & (72\%) \\
        7 & 0.125 & 10 & $100$M & $+0.00$ & 0.0005 & $-0.15$ & $-0.16$ & (84\%) & $-0.19$ & (90\%) \\
        7 & 0.125 & 10 & $300$M & $-0.00$ & 0.0002 & $-0.17$ & $-0.17$ & (66\%) & $-0.17$ & (88\%) \\
        \hline
        \multicolumn{11}{c}{\emph{$N=7$, $r_s=1$, $\theta=1.0$ (High density, warm)}} \\
        \hline
        7 & 1.0 & 1 & $1$M & $+0.41$ & 0.0243 & $+7.68$ & $+8.79$ & (80\%) & $+8.07$ & (90\%) \\
        7 & 1.0 & 1 & $3$M & $+0.31$ & 0.0070 & $+8.31$ & $+8.65$ & (93\%) & $+8.65$ & (97\%) \\
        7 & 1.0 & 1 & $10$M & $+0.29$ & 0.0056 & $+8.34$ & $+8.64$ & (98\%) & $+8.46$ & (99\%) \\
        7 & 1.0 & 1 & $30$M & $+0.32$ & 0.0086 & $+8.33$ & $+8.64$ & (99\%) & $+8.44$ & (100\%) \\
        7 & 1.0 & 1 & $100$M & $+0.31$ & 0.0065 & $+8.42$ & $+8.56$ & (100\%) & $+8.44$ & (100\%) \\
        7 & 1.0 & 1 & $300$M & $+0.31$ & 0.0074 & $+8.40$ & $+8.53$ & (100\%) & $+8.39$ & (100\%) \\
        \hline
        \multicolumn{11}{c}{\emph{$N=7$, $r_s=1$, $\theta=0.125$ (High density, cold)}} \\
        \hline
        7 & 0.125 & 1 & $1$M & $+0.01$ & 0.0016 & $+3.83$ & $-1.06$ & (1\%) & $+3.78$ & (83\%) \\
        7 & 0.125 & 1 & $3$M & $+0.00$ & 0.0003 & $-9.05$ & $-2.09$ & (1\%) & $-22.50$ & (66\%) \\
        7 & 0.125 & 1 & $10$M & $+0.01$ & 0.0004 & $-0.72$ & $-1.16$ & (7\%) & $-1.65$ & (26\%) \\
        7 & 0.125 & 1 & $30$M & $+0.00$ & 0.0002 & $-9.41$ & $-1.51$ & (2\%) & $-5.71$ & (79\%) \\
        7 & 0.125 & 1 & $100$M & $+0.00$ & 0.0000 & $-83.17$ & $-0.76$ & (67\%) & $-4.90$ & (79\%) \\
        7 & 0.125 & 1 & $300$M & $-0.00$ & 0.0001 & $-2.62$ & $-0.91$ & (47\%) & $-0.39$ & (88\%) \\
        \hline\hline
    \end{tabular}
    \caption{As Table~\ref{tab:ueg-convergence-n19}, for $N=7$ ($p(7)=15$ families).}
    \label{tab:ueg-convergence-n7}
\end{table*}

\begin{table*}[htbp]
    \centering
    \begin{tabular}{c c c r | r | r | r | r c | r c}
        \hline\hline
        $N$ & $\theta$ & $r_s$ & Steps & $\bar{\sigma}$ & $\text{KL}_{\text{MAP}}$ & $E_{\text{MC}}/N$ & \multicolumn{2}{c|}{$E_{\text{Lin}}^{\text{MAP}}/N$ ($R^2$)} & \multicolumn{2}{c}{$E_{\text{Deep}}^{\text{LSTM}}/N$ ($R^2$)} \\
        \hline
        \multicolumn{11}{c}{\emph{$N=33$, $r_s=10$, $\theta=1.0$ (Low density, warm)}} \\
        \hline
        33 & 1.0 & 10 & $1$M & $+0.35$ & 0.0075 & $-0.04$ & $-0.05$ & (36\%) & $-0.05$ & (84\%) \\
        33 & 1.0 & 10 & $3$M & $+0.43$ & 0.0033 & $-0.05$ & $-0.05$ & (88\%) & $-0.05$ & (97\%) \\
        33 & 1.0 & 10 & $10$M & $+0.37$ & 0.0048 & $-0.04$ & $-0.05$ & (78\%) & $-0.05$ & (75\%) \\
        33 & 1.0 & 10 & $30$M & $+0.39$ & 0.0019 & $-0.05$ & $-0.05$ & (81\%) & $-0.05$ & (91\%) \\
        33 & 1.0 & 10 & $100$M & $+0.40$ & 0.0025 & $-0.05$ & $-0.05$ & (93\%) & $-0.05$ & (95\%) \\
        33 & 1.0 & 10 & $300$M & $+0.40$ & 0.0016 & $-0.05$ & $-0.05$ & (99\%) & $-0.05$ & (98\%) \\
        \hline
        \multicolumn{11}{c}{\emph{$N=33$, $r_s=10$, $\theta=0.125$ (Low density, cold)}} \\
        \hline
        33 & 0.125 & 10 & $1$M & $-0.08$ & 1.9717 & $-0.19$ & $-0.19$ & (4\%) & $-0.21$ & (84\%) \\
        33 & 0.125 & 10 & $3$M & $-0.05$ & 1.3151 & $-0.18$ & $-0.19$ & (2\%) & $-0.20$ & (75\%) \\
        33 & 0.125 & 10 & $10$M & $+0.00$ & 0.8808 & $-0.25$ & $-0.19$ & (4\%) & $-0.19$ & (82\%) \\
        33 & 0.125 & 10 & $30$M & $-0.01$ & 0.5953 & $-0.19$ & $-0.21$ & (11\%) & $-0.20$ & (53\%) \\
        33 & 0.125 & 10 & $100$M & $-0.00$ & 0.4416 & $-0.66$ & $-0.20$ & (18\%) & $-0.20$ & (41\%) \\
        33 & 0.125 & 10 & $300$M & $-0.00$ & 0.3968 & $-0.21$ & $-0.22$ & (29\%) & $-0.20$ & (21\%) \\
        \hline
        \multicolumn{11}{c}{\emph{$N=33$, $r_s=1$, $\theta=1.0$ (High density, warm)}} \\
        \hline
        33 & 1.0 & 1 & $1$M & $+0.07$ & 0.0764 & $+7.67$ & $+8.52$ & (35\%) & $+10.39$ & (26\%) \\
        33 & 1.0 & 1 & $3$M & $-0.03$ & 0.0241 & $+6.20$ & $+8.73$ & (48\%) & $+5.72$ & (19\%) \\
        33 & 1.0 & 1 & $10$M & $-0.02$ & 0.0140 & $+6.25$ & $+8.49$ & (69\%) & $+6.71$ & (33\%) \\
        33 & 1.0 & 1 & $30$M & $+0.01$ & 0.0146 & $+9.05$ & $+8.82$ & (82\%) & $-96.29$ & (46\%) \\
        33 & 1.0 & 1 & $100$M & $+0.00$ & 0.0131 & $-1.03$ & $+8.72$ & (92\%) & $+7.90$ & (65\%) \\
        33 & 1.0 & 1 & $300$M & $+0.00$ & 0.0092 & $+9.86$ & $+8.74$ & (95\%) & $+8.00$ & (77\%) \\
        \hline
        \multicolumn{11}{c}{\emph{$N=33$, $r_s=1$, $\theta=0.125$ (High density, cold)}} \\
        \hline
        33 & 0.125 & 1 & $1$M & $-0.00$ & 1.0456 & $-6.11$ & $-1.27$ & (3\%) & $-5.58$ & (52\%) \\
        33 & 0.125 & 1 & $3$M & $-0.03$ & 0.6395 & $-2.42$ & $-1.89$ & (1\%) & $-0.94$ & (22\%) \\
        33 & 0.125 & 1 & $10$M & $-0.00$ & 0.3803 & $-3.24$ & $-1.71$ & (0\%) & $-0.15$ & (5\%) \\
        33 & 0.125 & 1 & $30$M & $-0.00$ & 0.2795 & $-2.87$ & $-1.81$ & (2\%) & $-1.65$ & (6\%) \\
        33 & 0.125 & 1 & $100$M & $+0.00$ & 0.2361 & $-2.31$ & $-1.80$ & (4\%) & $-1.52$ & (11\%) \\
        33 & 0.125 & 1 & $300$M & $+0.00$ & 0.2157 & $-2.01$ & $-1.82$ & (11\%) & $-1.82$ & (21\%) \\
        \hline\hline
    \end{tabular}
    \caption{As Table~\ref{tab:ueg-convergence-n19}, for $N=33$ ($p(33)=10\,143$ families). DuBois references~\cite{dubois2017overcoming}: $+17.38 \, \text{Ry}/N$ ($r_s=1$, $\theta=1$) and $-0.208 \, \text{Ry}/N$ ($r_s=10$, $\theta=0.125$).}
    \label{tab:ueg-convergence-n33}
\end{table*}

\bibliography{Halcyon-PermutationFamily}

@misc{Halcyon.jl,
  title        = {{Halcyon.jl}: PIGS might fly},
  author       = {Frost, Jarvist Moore},
  year         = {2026},
  howpublished = {\url{https://github.com/frost-group/Halcyon.jl}},
  note         = {Julia package, Commit d31d95e}
}

@article{dubois2017overcoming,
  title={Overcoming the fermion sign problem in homogeneous systems},
  author={DuBois, Jonathan L and Brown, Ethan W and Alder, Berni J},
  journal={Advances in the Computational Sciences},
  pages={184--192},
  year={2017},
  publisher={World Scientific},
  
  eprint={1409.3262},
  archivePrefix = {arXiv},
  primaryClass  = {cond-mat.str-el},

  url={https://arxiv.org/abs/1409.3262},
  doi = {10.48550/ARXIV.1409.3262}
}

@phdthesis{brown2014thesis-PATHINTEGRALMONTE,
  type = {Thesis},
  title = {{{PATH INTEGRAL MONTE CARLO AND THE ELECTRON GAS}}},
  author = {Brown, Ethan},
  year = 2014,
  school = {University of Illinois at Urbana-Champaign},
}

@article{Ceperley1995,
  title = {Path integrals in the theory of condensed helium},
  volume = {67},
  ISSN = {1539-0756},
  url = {http://dx.doi.org/10.1103/RevModPhys.67.279},
  DOI = {10.1103/revmodphys.67.279},
  number = {2},
  journal = {Reviews of Modern Physics},
  publisher = {American Physical Society (APS)},
  author = {Ceperley,  D. M.},
  year = {1995},
  month = Apr,
  pages = {279–355}
}

@article{Hochreiter1997,
  title = {Long Short-Term Memory},
  volume = {9},
  ISSN = {1530-888X},
  url = {http://dx.doi.org/10.1162/neco.1997.9.8.1735},
  DOI = {10.1162/neco.1997.9.8.1735},
  number = {8},
  journal = {Neural Computation},
  publisher = {MIT Press},
  author = {Hochreiter,  Sepp and Schmidhuber,  J\"{u}rgen},
  year = {1997},
  month = Nov,
  pages = {1735–1780}
}

@article{Dornheim2019permutation,
  title = {Path integral Monte Carlo simulation of degenerate electrons: Permutation-cycle properties},
  volume = {151},
  ISSN = {1089-7690},
  url = {http://dx.doi.org/10.1063/1.5093171},
  DOI = {10.1063/1.5093171},
  number = {1},
  journal = {The Journal of Chemical Physics},
  publisher = {AIP Publishing},
  author = {Dornheim,  T. and Groth,  S. and Filinov,  A. V. and Bonitz,  M.},
  year = {2019},
  month = jul 
}

@article{Dornheim2019-UEG-2DTRAP,
  title = {Fermion sign problem in path integral Monte Carlo simulations: Quantum dots,  ultracold atoms,  and warm dense matter},
  volume = {100},
  ISSN = {2470-0053},
  url = {http://dx.doi.org/10.1103/PhysRevE.100.023307},
  DOI = {10.1103/physreve.100.023307},
  number = {2},
  journal = {Physical Review E},
  publisher = {American Physical Society (APS)},
  author = {Dornheim,  T.},
  year = {2019},
  month = aug 
}

@article{Feynman1953A,
  title = {Atomic Theory of Liquid Helium Near Absolute Zero},
  volume = {91},
  ISSN = {0031-899X},
  url = {http://dx.doi.org/10.1103/PhysRev.91.1301},
  DOI = {10.1103/physrev.91.1301},
  number = {6},
  journal = {Physical Review},
  publisher = {American Physical Society (APS)},
  author = {Feynman,  R. P.},
  year = {1953},
  month = sep,
  pages = {1301–1308}
}

@article{Feynman1953B,
  title = {Atomic Theory of the lambda Transition in Helium},
  volume = {91},
  ISSN = {0031-899X},
  url = {http://dx.doi.org/10.1103/PhysRev.91.1291},
  DOI = {10.1103/physrev.91.1291},
  number = {6},
  journal = {Physical Review},
  publisher = {American Physical Society (APS)},
  author = {Feynman,  R. P.},
  year = {1953},
  month = sep,
  pages = {1291–1301}
}

@article{bezanson2017julia,
  title={Julia: A fresh approach to numerical computing},
  author={Bezanson, Jeff and Edelman, Alan and Karpinski, Stefan and Shah, Viral B},
  journal={SIAM review},
  volume={59},
  number={1},
  pages={65--98},
  year={2017},
  publisher={SIAM},
  url={https://doi.org/10.1137/141000671}
}

@article{Spada2022,
  title = {Path-Integral Monte Carlo Worm Algorithm for Bose Systems with Periodic Boundary Conditions},
  volume = {7},
  ISSN = {2410-3896},
  url = {http://dx.doi.org/10.3390/condmat7020030},
  DOI = {10.3390/condmat7020030},
  number = {2},
  journal = {Condensed Matter},
  publisher = {MDPI AG},
  author = {Spada,  Gabriele and Giorgini,  Stefano and Pilati,  Sebastiano},
  year = {2022},
  month = mar,
  pages = {30}
}

@article{Xiong2022,
  title = {On the thermodynamic properties of fictitious identical particles and the application to fermion sign problem},
  volume = {157},
  ISSN = {1089-7690},
  url = {http://dx.doi.org/10.1063/5.0106067},
  DOI = {10.1063/5.0106067},
  number = {9},
  journal = {The Journal of Chemical Physics},
  publisher = {AIP Publishing},
  author = {Xiong,  Yunuo and Xiong,  Hongwei},
  year = {2022},
  month = sep 
}

@article{Hennig2015,
  title = {Probabilistic numerics and uncertainty in computations},
  volume = {471},
  ISSN = {1471-2946},
  url = {http://dx.doi.org/10.1098/rspa.2015.0142},
  DOI = {10.1098/rspa.2015.0142},
  number = {2179},
  journal = {Proceedings of the Royal Society A: Mathematical,  Physical and Engineering Sciences},
  publisher = {The Royal Society},
  author = {Hennig,  Philipp and Osborne,  Michael A. and Girolami,  Mark},
  year = {2015},
  month = jul,
  pages = {20150142}
}

@article{Wang2001,
  title = {Efficient,  Multiple-Range Random Walk Algorithm to Calculate the Density of States},
  volume = {86},
  ISSN = {1079-7114},
  url = {http://dx.doi.org/10.1103/PhysRevLett.86.2050},
  DOI = {10.1103/physrevlett.86.2050},
  number = {10},
  journal = {Physical Review Letters},
  publisher = {American Physical Society (APS)},
  author = {Wang,  Fugao and Landau,  D. P.},
  year = {2001},
  month = mar,
  pages = {2050–2053}
}

@article{Inglis2013,
  title = {Wang-Landau method for calculating Rényi entropies in finite-temperature quantum Monte Carlo simulations},
  volume = {87},
  ISSN = {1550-2376},
  url = {http://dx.doi.org/10.1103/PhysRevE.87.013306},
  DOI = {10.1103/physreve.87.013306},
  number = {1},
  journal = {Physical Review E},
  publisher = {American Physical Society (APS)},
  author = {Inglis,  Stephen and Melko,  Roger G.},
  year = {2013},
  month = jan 
}

@article{Miller1964,
  title = {A Trustworthy Jackknife},
  volume = {35},
  ISSN = {0003-4851},
  url = {http://dx.doi.org/10.1214/aoms/1177700384},
  DOI = {10.1214/aoms/1177700384},
  number = {4},
  journal = {The Annals of Mathematical Statistics},
  publisher = {Institute of Mathematical Statistics},
  author = {Miller,  Rupert G.},
  year = {1964},
  month = dec,
  pages = {1594–1605}
}

@article{Miller1974,
  title = {The Jackknife--A Review},
  volume = {61},
  ISSN = {0006-3444},
  url = {http://dx.doi.org/10.2307/2334280},
  DOI = {10.2307/2334280},
  number = {1},
  journal = {Biometrika},
  publisher = {JSTOR},
  author = {Miller,  Rupert G.},
  year = {1974},
  month = apr,
  pages = {1}
}

@article{Boninsegni2006A,
  title = {Worm Algorithm for Continuous-Space Path Integral Monte Carlo Simulations},
  volume = {96},
  ISSN = {1079-7114},
  url = {http://dx.doi.org/10.1103/PhysRevLett.96.070601},
  DOI = {10.1103/physrevlett.96.070601},
  number = {7},
  journal = {Physical Review Letters},
  publisher = {American Physical Society (APS)},
  author = {Boninsegni,  Massimo and Prokof’ev,  Nikolay and Svistunov,  Boris},
  year = {2006},
  month = feb 
}

@article{Boninsegni2006B,
  title = {Worm algorithm and diagrammatic Monte Carlo: A new approach to continuous-space path integral Monte Carlo simulations},
  volume = {74},
  ISSN = {1550-2376},
  url = {http://dx.doi.org/10.1103/PhysRevE.74.036701},
  DOI = {10.1103/physreve.74.036701},
  number = {3},
  journal = {Physical Review E},
  publisher = {American Physical Society (APS)},
  author = {Boninsegni,  M. and Prokof’ev,  N. V. and Svistunov,  B. V.},
  year = {2006},
  month = sep
}

@article{Hermann2023,
  title = {Ab initio quantum chemistry with neural-network wavefunctions},
  volume = {7},
  ISSN = {2397-3358},
  url = {http://dx.doi.org/10.1038/s41570-023-00516-8},
  DOI = {10.1038/s41570-023-00516-8},
  number = {10},
  journal = {Nature Reviews Chemistry},
  publisher = {Springer Science and Business Media LLC},
  author = {Hermann,  Jan and Spencer,  James and Choo,  Kenny and Mezzacapo,  Antonio and Foulkes,  W. M. C. and Pfau,  David and Carleo,  Giuseppe and Noé,  Frank},
  year = {2023},
  month = aug,
  pages = {692–709}
}

@article{Filinov2003,
  title = {Improved Kelbg potential for correlated Coulomb systems},
  volume = {36},
  ISSN = {1361-6447},
  url = {http://dx.doi.org/10.1088/0305-4470/36/22/317},
  DOI = {10.1088/0305-4470/36/22/317},
  number = {22},
  journal = {Journal of Physics A: Mathematical and General},
  publisher = {IOP Publishing},
  author = {Filinov,  A V and Bonitz,  M and Ebeling,  W},
  year = {2003},
  month = May,
  pages = {5957–5962}
}

@article{Yakub2003,
  title = {An efficient method for computation of long-ranged Coulomb forces in computer simulation of ionic fluids},
  volume = {119},
  ISSN = {1089-7690},
  url = {http://dx.doi.org/10.1063/1.1624364},
  DOI = {10.1063/1.1624364},
  number = {22},
  journal = {The Journal of Chemical Physics},
  publisher = {AIP Publishing},
  author = {Yakub,  Eugene and Ronchi,  Claudio},
  year = {2003},
  month = Dec,
  pages = {11556–11560}
}

@article{Troyer2005,
  title = {Computational Complexity and Fundamental Limitations to Fermionic Quantum Monte Carlo Simulations},
  volume = {94},
  ISSN = {1079-7114},
  url = {http://dx.doi.org/10.1103/PhysRevLett.94.170201},
  DOI = {10.1103/physrevlett.94.170201},
  number = {17},
  journal = {Physical Review Letters},
  publisher = {American Physical Society (APS)},
  author = {Troyer,  Matthias and Wiese,  Uwe-Jens},
  year = {2005},
  month = May 
}

@article{dornheimPathIntegralMonte2019,
    title = {Path integral {Monte} {Carlo} simulation of degenerate electrons: {Permutation}-cycle properties},
    volume = {151},
    issn = {0021-9606},
    shorttitle = {Path integral {Monte} {Carlo} simulation of degenerate electrons},
    url = {https://aip.scitation.org/doi/abs/10.1063/1.5093171},
    doi = {10.1063/1.5093171},
    number = {1},
    urldate = {2020-02-16},
    journal = {The Journal of Chemical Physics},
    author = {Dornheim, T. and Groth, S. and Filinov, A. V. and Bonitz, M.},
    month = jul,
    year = {2019},
    pages = {014108},
}

@misc{hpc,
  title = {{Research Computing Service}},
  author = {{Imperial College London}},
  doi = {10.14469/hpc/2232}
}

\end{document}